\documentclass[a4paper]{article}
\usepackage{pdfsync}
\usepackage[utf8]{inputenc}
\usepackage{graphicx}
\usepackage{amssymb,amsfonts,amsmath,amsthm}
\usepackage{hyperref}
\usepackage{graphicx}
\usepackage{multirow}
\usepackage{verbatim}
\usepackage{color}
\usepackage{booktabs}
\usepackage[sectionbib,round]{natbib}
\usepackage{textcomp}
\usepackage{upquote}

\theoremstyle{plain}

\theoremstyle{definition}

\newtheorem{remark}{Remark}[section]

\begin{document}

\title{\textbf{wavScalogram: an R package with wavelet scalogram tools for time series analysis}}


\author{V. J. Bol\'os$^1$, R. Ben\'{\i}tez$^1$ \\ \\
{\small $^1$ Dpto. Matem\'aticas para la Econom\'{\i}a y la Empresa, Facultad de Econom\'{\i}a,} \\
{\small Universidad de Valencia. Avda. Tarongers s/n, 46022 Valencia, Spain.} \\
{\small e-mail\textup{: \texttt{vicente.bolos@uv.es} (V. J. Bol\'os), \texttt{rabesua@uv.es} (R. Ben\'{\i}tez)}} \\}

\date{June 2022}

\maketitle

\begin{abstract}
In this work we present the \textbf{wavScalogram} R package, which contains methods based on wavelet scalograms for time series analysis. These methods are related to two main wavelet tools: the windowed scalogram difference and the scale index. The windowed scalogram difference compares two time series, identifying if their scalograms follow similar patterns at different scales and times, and it is thus a useful complement to other comparison tools such as the squared wavelet coherence. On the other hand, the scale index provides a numerical estimation of the degree of non-periodicity of a time series and it is widely used in many scientific areas. 
\end{abstract}

\section{Introduction}
Since the works of \cite{mal98} and \cite{dau92}, wavelet analysis has become, in the last  few decades, a standard tool in the  field of time series analysis. 
Its ability to simultaneously analyze a signal in frequency space (scales) and in time, allows it to overcome many of the limitations that Fourier analysis presents for non-stationary time series. Furthermore, the algorithms for calculating the different wavelet transforms are characterized by their speed and ease of implementation.

There are currently many software packages that implement functions for wavelet analysis of time series (MATLAB's  Wavelet Toolbox, Wavelab, etc.), and in recent years, the exponential growth of the R ecosystem has not been outside the field of wavelet analysis. Within CRAN there are many packages related to wavelet analysis for time series. Specifically, as collected in the \textbf{TimeSeries} Task View, the \textbf{wavelets} package (\cite{wavelets}) , the \textbf{WaveletComp} and \textbf{biwavelet} packages (\cite{WaveletComp, biwavelet}) , the \textbf{mvLSW} package (\cite{mvLSW}) and other packages such as \textbf{hwwntest} (\cite{hwwntest}), \textbf{rwt} (\cite{rwt}),  \textbf{waveslim} (\cite{waveslim}) and \textbf{wavethresh} (\cite{wavethresh}).

In this work we will describe in depth the \textbf{wavScalogram}  package (\cite{wavScalogram}) (also mentioned in the \textbf{TimeSeries} Task View). In this package, methods based on the wavelet scalogram are introduced as defined in \cite{ben10,bol17,bol20}. These methods are basically related to two main wavelet tools: the \textit{windowed scalogram difference} and the \textit{scale index}. The first one, the windowed scalogram difference, was introduced in \cite{bol17}. It allows to compare two time series at different scales and times, determining if their scalograms follow similar patterns. In this sense, it is a complement to other wavelet tools for comparing time series such as the squared wavelet coherence and the phase difference, since there are certain  differences in time series that these measurements are not capable of detecting while the windowed scalogram difference can. The second tool is the scale index introduced in \cite{ben10}. It focuses on the analysis of the non-periodicity of a signal, giving a numerical measure of its degree of non-periodicity, taking the value $0$ if the signal is periodic and a value close to $1$ if the signal is totally aperiodic (for example, a purely stochastic signal). The scale index has been used in many scientific areas, being the evaluation of the quality of pseudo-random number generators the area where it has been used the most. In addition, the scale index also has a \textit{``windowed''} version, in which the windowed scalogram is used to calculate the scale index instead, allowing to measure the evolution of the scale index over time, which is useful in the case of non-stationary time series (see \cite{bol20}).

The article is organized as follows: In the next section, we describe the basics of the wavelet analysis and how to use them in the \textbf{wavScalogram} package. Then,  a description of the wavelet scalogram and its implementation is given. The following sections are devoted to the windowed scalogram difference and the scale index, in its original and windowed versions. Finally we illustrate the use of the package with some examples in applied problems, such as the analysis of time series of sunspots or the use of the windowed scalogram difference in the clustering of time series, particularly the interest rate series of sovereign bonds.

\section{Wavelet introduction}
\label{sec2}

A \textit{wavelet} (or \textit{mother wavelet}) is a function $\psi \in L^2\left( \mathbb{R}\right) $ with zero average (i.e.  $\int _{\mathbb{R}} \psi =0$), unit energy ($\| \psi \| =1$, i.e. normalized) and centered in the neighborhood of $t=0$ (\cite{mal98}). There exists a wide variety of wavelets but in this package we use the following, described in \cite{tor98} (see Figure~\ref{fig:wav}):

\begin{figure}[tbp]
\begin{center}
  \includegraphics[width=1\textwidth]{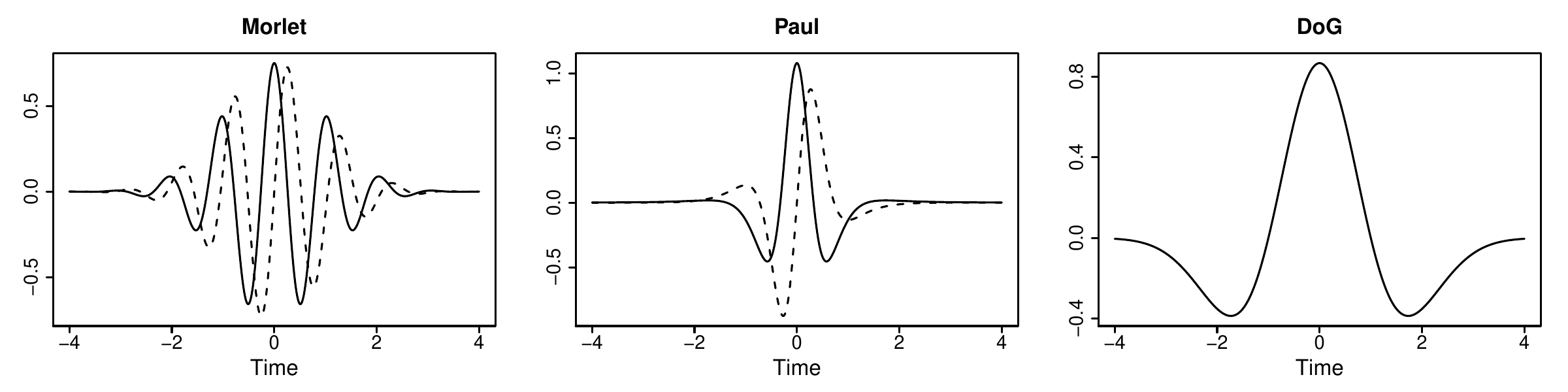}
\end{center}
\caption{Real part (solid) and imaginary part (dashed) of Morlet, Paul and DoG wavelets for default parameter values, $\omega _0=6$ and $m=4,2$ respectively. }
\label{fig:wav}
\end{figure}

\begin{itemize}
\item Morlet:
\[
\psi _{\textrm{Morlet}}(t)=\pi^{-1/4}e^{\mathrm{i}\omega_0 t}e^{-t^2/2}.
\]
It is a plane wave modulated by a Gaussian, where the positive parameter $\omega _0$ denotes the central dimensionless frequency. According to \cite{far92}, the wavelet function must fulfil an admissibility condition, which for the Morlet wavelet is only accomplished if some correction factors are added. We take as default value $\omega _0=6$, for which those correction factors are negligible. Nevertheless, other choices of this parameter can be considered.
\item Paul:
\[
\psi _{\textrm{Paul}}(t)=\frac{\left(2\mathrm{i}\right)^m m!}{\sqrt{\pi\left(2m\right)!}}\left( 1-\mathrm{i}t\right)^{-\left( m+1\right)},
\]
where $m$ is a positive integer parameter representing the order. By default, $m=4$.
\item Derivative of a Gaussian (DoG):
\[
\psi _{\textrm{DoG}}(t)=\frac{\left(-1\right)^{m+1}}{\sqrt{\Gamma\left( m+\frac{1}{2}\right)}}\frac{\mathrm{d}^m}{\mathrm{d}t^m}\left( e^{-t^2/2}\right) ,
\]
where $m$ is a positive integer parameter representing the derivative. By default, $m=2$, that coincides with the Marr or Mexican hat wavelet.
\end{itemize}

Moreover, we have added:
\begin{itemize}
\item Haar, centered at $0$:
\[
\psi _{\textrm{Haar}}(t)=\left\{
\begin{array}{rl}
1 & \textrm{if }\,-\frac{1}{2}\leq t <0, \\
-1 & \textrm{if }\,0\leq t <\frac{1}{2}, \\
0 & \textrm{otherwise}.
\end{array}
\right.
\]
This is the simplest wavelet, but it is not continuous.
\end{itemize}

Scaling a wavelet $\psi $ by $s>0$ and translating it by $u$, we create a family of unit energy ``time-frequency atoms'', called \textit{daughter wavelets}, $\psi _{u,s}$, as follows
\begin{equation}
\label{eq:psius}
\psi_{u,s}(t)=\frac{1}{\sqrt{s}}\psi \left( \frac{t-u}{s}\right) .
\end{equation}

\begin{remark}[Fourier factor]
\label{rem:ff}
Usually, the Fourier wavelength of a daughter wavelet does not coincide with its scale $s$. Nevertheless, they are proportional, and this proportionality factor for converting scales into Fourier periods is called \textit{Fourier factor}. This Fourier factor is taken $4\pi/\left( \omega _0+\sqrt{2+\omega_0^2}\right) $, $4\pi/\left( 2m+1\right) $ and $2\pi / \sqrt{m+1/2}$ for Morlet, Paul and DoG wavelets respectively (\cite{tor98}). For the default parameter values, the Fourier factor is approximately $1.033$, $1.3963$, and $3.9738$ respectively. For Haar wavelet, the Fourier factor is $1$.
\end{remark}

Given a function $f\in L^2\left( \mathbb{R}\right) $, that we will identify with a \textit{signal} or \textit{time series}, the \textit{continuous wavelet transform} (CWT) of $f$ at time $u$ and scale $s>0$ is defined as
\begin{equation}
\label{eq:cwt}
\mathcal{W}f\left( u,s\right) =\int _{-\infty}^{+\infty}f(t)\psi_{u,s}^* (t) \, \textrm{d}t,
\end{equation}
where $ ^*$ denotes the complex conjugate. The CWT allows us to obtain the frequency components (or \textit{details}) of $f$ corresponding to scale $s$ and time location $u$.

In practical situations, however, it is common to deal with finite signals. That is, given a time signal $x$, and a finite time interval $[0,T]$, we shall consider the finite sequence $x_n=x\left(t_n\right)$, for $n=0,\ldots,N$. Here, $t_0,\ldots,t_{N}$ is a discretization of the interval $[0,T]$, i.e. $t_n = nh$, being $h = T/N$ the time step.  According to \eqref{eq:psius} and \eqref{eq:cwt}, the CWT of $x$ at scale $s>0$ is defined as the sequence
\begin{equation}
\label{eq:wxns}
\mathcal{W}x_n(s)=h\sum _{i=0}^{N} x_i \psi _{n,s}^*\left( t_i\right) ,
\end{equation}
where $n=0,\ldots,N$ and
\begin{equation}
\label{eq:psins}
\psi_{n,s}(t)=\frac{1}{\sqrt{s}}\psi \left( \frac{t-t_n}{s}\right) .
\end{equation}
Note that $\psi_{n,s}(t)$ is in fact $\psi_{t_n,s}(t)$, but this abuse of notation between \eqref{eq:psius} and \eqref{eq:psins} is assumed for the sake of readability. Using Fourier transform tools, one can calculate \eqref{eq:wxns} for all $n=0,\ldots N$ simultaneously and efficiently (\cite{tor98}).

\begin{remark}[\textbf{Energy density}]
\label{rmk:cwt}
It is known that the CWT coefficients are biased in favour of large scales (\cite{liu07}). Nevertheless, if the mother and daughter wavelets are normalized by the $L^1$-norm (as in the \textbf{Rwave} package, by \cite{Rwave}) instead of the $L^2$-norm (as in our package), this bias is not produced . Hence, to rectify the bias, the CWT in \eqref{eq:cwt} and \eqref{eq:wxns} can be multiplied by the factor $\frac{1}{\sqrt{s}}$. This rectification will be specially useful in some wavelet tools of our package that quantify the ``energy density'' of a signal, such as the wavelet power spectrum, the scalograms and the windowed scalogram difference. On the other hand, in the case of the scale index, this correction will not be advisable (see Remark \ref{rmk:noed}). Usually, the wavelet tools of our package have a logical parameter called \texttt{energy\_density} that switches this correction.
\end{remark}

For computing the CWT of a time series $x$ at a given set of scales, we use \texttt{cwt\_wst}. For example:
\begin{verbatim}
> # install.packages("wavScalogram")
> library(wavScalogram)
> h <- 0.1
> N <- 1000
> time <- seq(from = 0, to = N * h, by = h)
> signal <- sin(pi * time)
> scales <- seq(from = 0.5, to = 4, by = 0.05)
> cwt <- cwt_wst(signal = signal, dt = h,
+                scales = scales, powerscales = FALSE,
+                wname = "DOG", wparam = 6)
\end{verbatim}
computes the CWT of \texttt{signal} at \texttt{scales} from $s_a=0.5$ to $s_b=4$ using DoG wavelet with $m=6$. The parameter \texttt{wname} indicates the wavelet used, and it can be \texttt{"MORLET"} (default value), \texttt{"PAUL"}, \texttt{"DOG"}, \texttt{"HAAR"} or \texttt{"HAAR2"}. The difference between these two last values is that  \texttt{"HAAR2"} provides a more accurate but slower algorithm than the one provided by \texttt{"HAAR"}. Moreover, we can specify by means of \texttt{wparam} the value of the parameters $\omega_0$ or $m$. As it has been stated before, the default values of these parameters are $\omega_0=6$ for Morlet wavelet, $m=4$ for Paul wavelet and $m=2$ for DoG wavelet.

If the set of scales is a base $2$ power scales set (\cite{tor98}), the parameter \texttt{scales} can be a vector of three elements with the lowest scale $s_a$, the highest scale $s_b$ and the number of suboctaves per octave. This vector is internally passed to function \texttt{pow2scales} that returns the constructed base $2$ power scales set. For example:
\begin{verbatim}
> scales <- c(0.5, 4, 16)
> cwt <- cwt_wst(signal = signal, dt = h, scales = scales,
+                powerscales = TRUE)
\end{verbatim}
computes the CWT of \texttt{signal} at scales from $s_a=0.5$ to $s_b=4$, with $16$ suboctaves per octave. Since parameter \texttt{powerscales} is \texttt{TRUE} by default, it is not necessary to specify it in the function call. If \texttt{scales = NULL} (default value), then the function constructs the scales set automatically: $s_a$ is chosen so that its equivalent Fourier period is $2h$ (\cite{tor98}), and $s_b=Nh/2r_w$, where $r_w$ is the corresponding \emph{wavelet radius}. Note that in this case $s_b$ maximizes the length of the scales interval taking into account the \emph{cone of influence}. The wavelet radius and the cone of influence are defined and discussed in Remark \ref{rem:coi}.

The output \texttt{cwt} is a list containing the following fields:

\begin{itemize}
\item \texttt{coefs} is an $(N+1)\times$\texttt{length(scales)} array (either real or complex depending on the wavelet used) containing the corresponding CWT coefficients, i.e. \texttt{cwt\$coefs[i,j]} is the CWT coefficient at the i-th time and j-th scale.
\item \texttt{scales} is the vector of scales used, either provided by the user or constructed by the function itself.
\item \texttt{fourier\_factor} is the scalar used to transform scales into Fourier periods (see Remark \ref{rem:ff}).
\item \texttt{coi\_maxscale} is a numeric vector of size $N+1$ that defines the \emph{cone of influence} (see Remark \ref{rem:coi}).
\end{itemize}

\begin{figure}[tbp]
\begin{center}
  \includegraphics[width=0.92\textwidth]{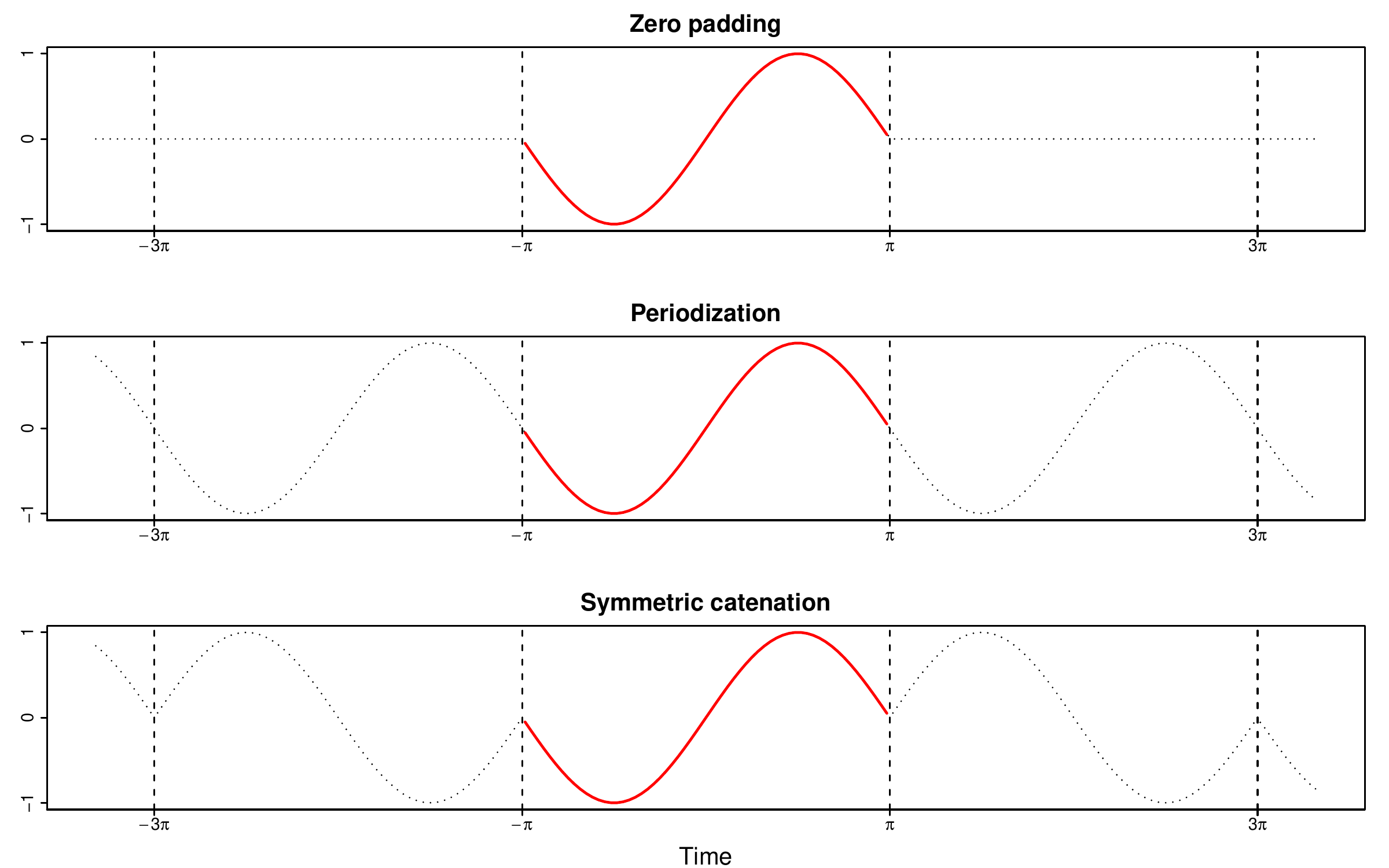}
\end{center}
\caption{Different constructions of an infinite signal from a finite length signal $\sin (t)$ with $t\in \left[-\pi ,\pi\right] $ (in red).}
\label{fig:be}
\end{figure}

\begin{remark}[Border effects]
\label{rem:be}
In \eqref{eq:wxns} (or \eqref{eq:cwt} for finite length signals) there appear \textit{border effects} (or \textit{edge effects}) when the support of the daughter wavelets is not entirely contained in the time domain $\left[ t_0,t_N\right] $. In order to try to mitigate border effects, we can construct from the original time series $x$ an infinite time series $\bar{x}$ on $t_i=t_0+ih$ for $i\in \mathbb{Z}$ and then we define
\begin{equation}
\label{eq:wbarxns}
\bar{\mathcal{W}}x_n(s)=\mathcal{W}\bar{x}_n(s)=h\sum _{i\in \mathbb{Z}} \bar{x}_i \psi _{n,s}^*\left( t_i\right) ,
\end{equation}
where $n=0,\ldots ,N$. The most usual ways to construct $\bar{x}$ are the following:
\begin{itemize}
\item Padding time series with zeroes: $\bar{x}_i=x_i$ if $i\in \left\{ 0,\ldots ,N\right\} $, and $\bar{x}_i=0$ otherwise. In this case, \eqref{eq:wxns} and \eqref{eq:wbarxns} are equivalent, having $\bar{\mathcal{W}}x_n(s)=\mathcal{W}x_n(s)$.
\item Using a periodization of the original time series: $\bar{x}_i=x_{i \,\textrm{mod}\,(N+1)}$.
\item Using a symmetric catenation of the original time series: $\bar{x}_i=x_{i \,\textrm{mod}\,(N+1)}$ if $\lfloor \frac{i}{N+1}\rfloor $ is even, and $\bar{x}_i=x_{(N-i) \,\textrm{mod}\,(N+1)}$ if $\lfloor \frac{i}{N+1}\rfloor $ is odd.
\end{itemize}

Depending on the nature of $x$, it may be preferable to use one construction or another for minimizing the undesirable border effects (see Figure~\ref{fig:be}). For example, a periodization is advised for stationary short time series, and symmetric catenation for non-stationary short time series. On the other hand, for long time series, border effects are less important and then we can just pad with zeroes, i.e. use the original CWT given in \eqref{eq:wxns}.
\end{remark}

How the border effects are treated by function \texttt{cwt\_wst} is determined via the parameter \texttt{border\_effects}. Possible values for this parameter are \texttt{"BE"} (raw border effects, padding with zeroes), \texttt{"PER"} (periodization) and \texttt{"SYM"} (symmetric catenation), corresponding to the three options described above.

\begin{remark}[Cone of influence]
\label{rem:coi}
The \textit{cone of influence} (CoI) is defined by the scales for which border effects become important at each time. The field \texttt{coi\_maxscale} of the output in \texttt{cwt\_wst} function contains, for each time, the maximum scale at which border effects are negligible, and consequently determines the CoI.

In order to compute the CoI, we have to set a criterion to distinguish between relevant and negligible border effects. In \cite{tor98}, the CoI is defined by the $e$-folding time for the autocorrelation of wavelet power spectrum (see Section \ref{sec:sc}) at each scale $s$, and this $e$-folding time is chosen so that the wavelet power spectrum for a discontinuity at the edge drops by a factor $e^{-2}$. For Morlet and DoG wavelets, this $e$-folding time is $\sqrt{2}s$, and for Paul wavelets is $s/\sqrt{2}$.

For wavelets with symmetric modulus such as  Morlet, Paul and DoG, the $e$-folding time at $s=1$ is interpreted as a \textit{wavelet radius} $r_w$ that defines an \textit{effective support} $\left[ -r_w,r_w\right] $ for the mother wavelet. Therefore, the CoI is given by the scales from which the corresponding effective supports of the daughter wavelets $\left[ u-sr_w,u+sr_w\right] $ are not entirely contained in the time domain.
\end{remark}

The wavelet radius $r_w$ determines the CoI in the different functions of this package by means of the parameter \texttt{waverad}. If it is \texttt{NULL} (default value) we consider $r_w=\sqrt{2}$ for Morlet and DoG wavelets, and $r_w=1/\sqrt{2}$ for Paul wavelets, following \cite{tor98}. On the other hand, we take $r_w=0.5$ for Haar wavelet, i.e. we assume that its effective support is in fact its support. Nevertheless we can introduce a custom $r_w$ for any wavelet, allowing us in this way to adjust the importance of border effects in the construction of the CoI. For example:
\begin{verbatim}
> cwt <- cwt_wst(signal = signal, dt = h,
+                wname = "DOG", wparam = 6, waverad = 2)
\end{verbatim}
computes the CWT coefficients of \texttt{signal} for DoG wavelet with $m=6$. Here, the value of \texttt{cwt\$coi\_maxscale} is obtained assuming that the wavelet radius is $r_w=2$. Note that the value of \texttt{waverad} does not affect the computation of the CWT coefficients.

\section{Wavelet scalograms}
\label{sec:sc}

The \textit{wavelet power spectrum} of a signal $f\in L^2\left( \mathbb{R}\right)$ at time $u$ and scale $s>0$ is defined as
\begin{equation}
\label{eq:wpsfus}
\mathcal{WPS}f\left( u,s\right) = |\mathcal{W}f\left( u,s\right) |^2.
\end{equation}
Analogously to \eqref{eq:wpsfus}, the wavelet power spectrum of a time series $x$ at scale $s>0$ is given by the sequence
\begin{equation}
\label{eq:wpsxns}
\mathcal{WPS}x_n(s) = |\mathcal{W}x_n(s) |^2,
\end{equation}
where $n=0,\ldots,N$.

\begin{figure}[tbp]
\begin{center}
  \includegraphics[width=1\textwidth]{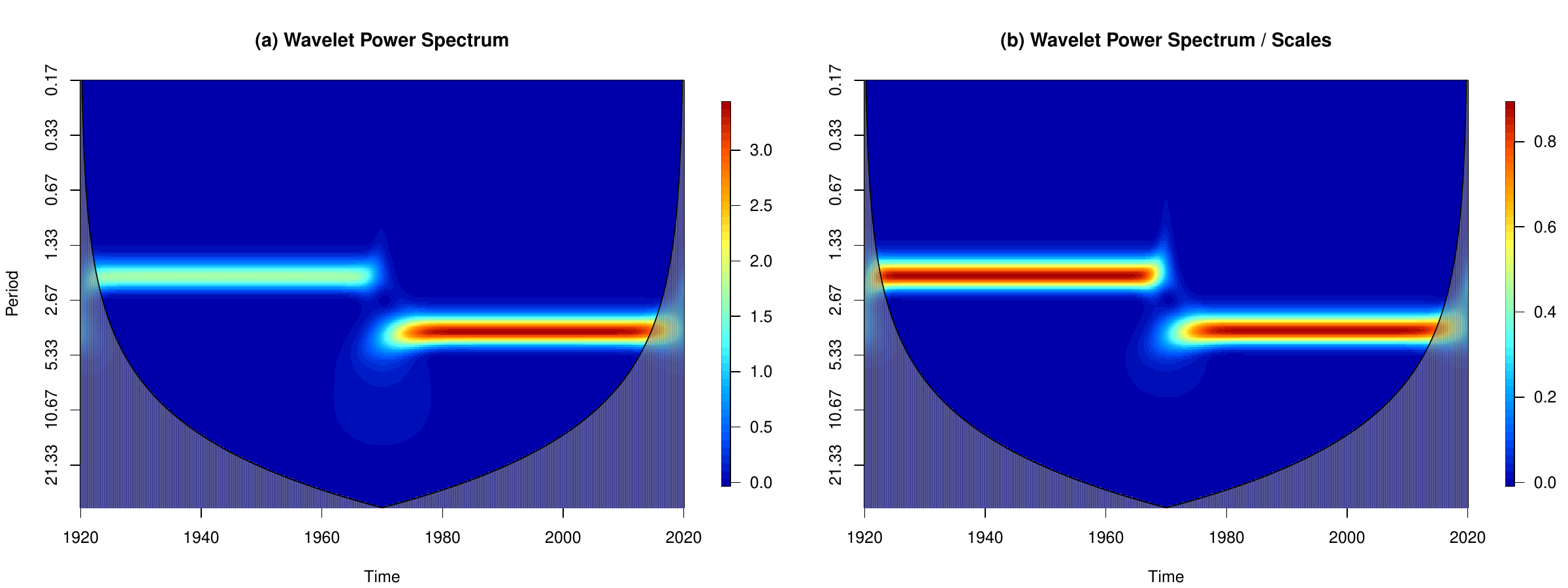}
\end{center}
\caption{Wavelet power spectra of \texttt{signal}, non corrected (a) and corrected (b) via parameter \texttt{energy\_density}. The CoI is the shadowed region. This signal is the concatenation of two pure sinusoidal time series with the same amplitude and different periods. Note that even though both time series have the same amplitude, when the coefficients are not corrected, the magnitude of the wavelet power spectrum is biased in favour of large scales, while in the corrected version, this bias is not present.}
\label{fig:wps}
\end{figure}

We can plot the wavelet power spectrum of a time series $x$ at a given set of scales through function \texttt{cwt\_wst} if the parameter \texttt{makefigure} is \texttt{TRUE} (default value). There are other parameters regarding this plot:
\begin{itemize}
\item \texttt{time\_values} is a vector that provides customized values in the time axis.
\item \texttt{energy\_density} is a logical parameter. If it is \texttt{TRUE}, it is plotted the wavelet power spectrum divided by the scales, according to Remark \ref{rmk:cwt}. By default, it is \texttt{FALSE}.
\item \texttt{figureperiod} is a logical parameter that indicates if they are represented periods or scales in the $y$-axis (see Remark \ref{rem:ff}). By default, it is \texttt{TRUE}.
\item \texttt{xlab, ylab, main, zlim} are parameters to customize the figure.
\end{itemize}
For example:
\begin{verbatim}
> h <- 1 / 12
> time1 <- seq(from = 1920, to = 1970 - h, by = h)
> time2 <- seq(from = 1970, to = 2020, by = h)
> signal <- c(sin(pi * time1), sin(pi * time2 / 2))
> cwt_a <- cwt_wst(signal = signal, dt = h, time_values = c(time1, time2))
> cwt_b <- cwt_wst(signal = signal, dt = h, time_values = c(time1, time2),
+                  energy_density = TRUE)
\end{verbatim}
plots Figure~\ref{fig:wps} (a) and (b) respectively. In this figure it is shown how the wavelet power spectrum is biased in favour of large scales, as it is pointed out in Remark \ref{rmk:cwt}. The parameter \texttt{energy\_density} corrects it and so, values for different scales become comparable. Note that \texttt{energy\_density} only affects the plot and hence, \texttt{cwt\_a} is identical to \texttt{cwt\_b}.

The \textit{scalogram} of $f$ at scale $s$ is defined as
\begin{equation}
\label{eq:sfs}
\mathcal{S}f(s)= \left( \int _{-\infty } ^{+\infty } | \mathcal{W}f\left( u,s\right) | ^2 \, \textrm{d}u \right) ^{1/2} .
\end{equation}
It gives the contribution of each scale to the total ``energy'' of the signal and so, the notion of scalogram here is analogous to the spectrum of the Fourier transform. It is important to note that the term ``scalogram'' is often used to refer the wavelet power spectrum, but in this package, we call ``scalogram'' to \eqref{eq:sfs}.

If $f$ is a finite length signal with time domain $I=\left[ a,b\right] $, it is usual to consider a normalized version of the scalogram for comparison purposes, given by
\begin{equation}
\label{eq:sfsn}
\mathcal{S}f(s)= \left( \frac{1}{b-a}\int _{a}^{b} | \mathcal{W}f\left( u,s\right) | ^2 \, \textrm{d}u \right) ^{1/2} .
\end{equation}
Hence, according to \eqref{eq:sfsn}, the \textit{(normalized) scalogram} of $x$ at scale $s$ is given by
\begin{equation}
\label{eq:sxsn}
\mathcal{S}x(s) = \left( \frac{1}{N+1}\sum _{i=0}^{N} | \mathcal{W}x_i(s) | ^2 \right) ^{1/2}.
\end{equation}
The normalization coefficient $1/(N+1)$ in \eqref{eq:sxsn} allows us to compare scalograms of time series with different lengths.

\begin{figure}[tbp]
\begin{center}
  \includegraphics[width=1\textwidth]{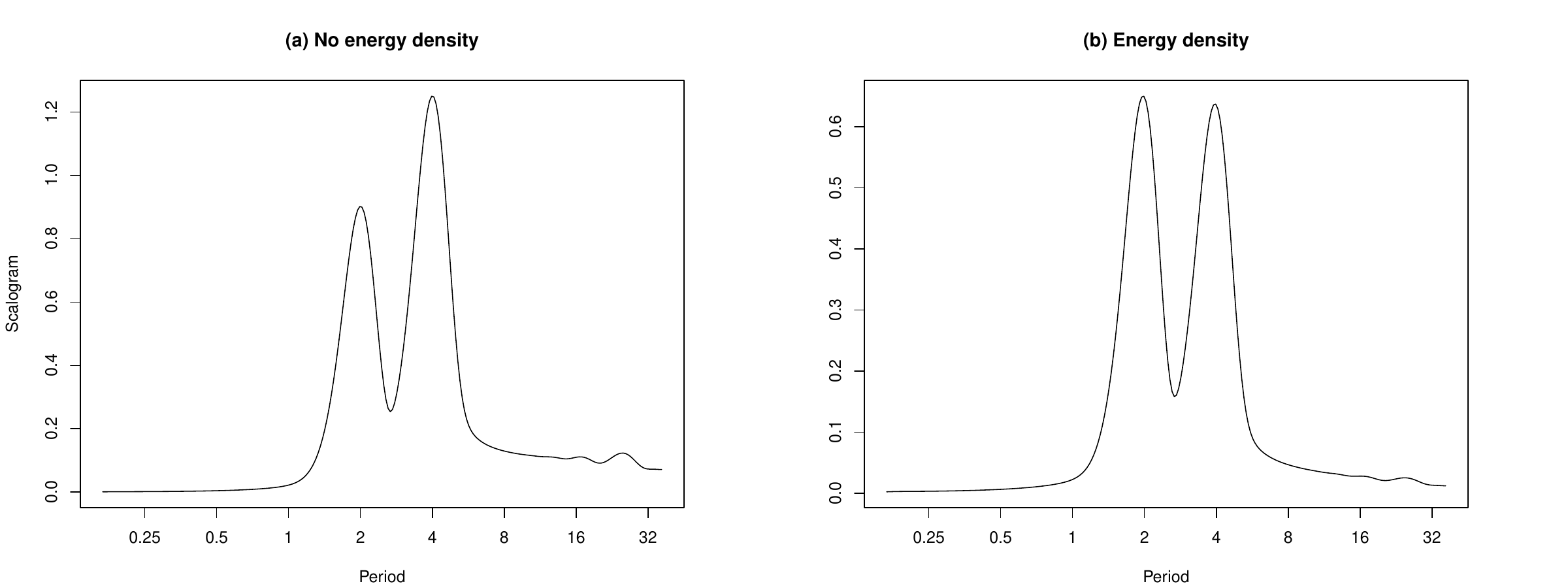}
\end{center}
\caption{Original scalogram (a) and corrected scalogram representing an energy density measure (b), both relative to \texttt{signal}. As in Figure \ref{fig:wps}, it can be seen how the scalogram is biased in favour of large scales when the parameter \texttt{energy\_density} is \texttt{FALSE} (plot (a)).} 
\label{fig:sc}
\end{figure}

We can compute the (normalized) scalograms of a time series $x$ at a given set of scales by means of function \texttt{scalogram}. This function follows the same rules as \texttt{cwt\_wst} regarding the data entry, construction of scales, choice of the wavelet, border effects and application of Fourier factor. So, parameters \texttt{signal}, \texttt{dt}, \texttt{scales}, \texttt{powerscales}, \texttt{wname}, \texttt{wparam}, \texttt{waverad}, \texttt{border\_effects}, \texttt{makefigure}, \texttt{figureperiod}, \texttt{xlab}, \texttt{ylab} and \texttt{main} are analogous to those in function \texttt{cwt\_wst} with same default values. On the other hand, parameter \texttt{energy\_density} is \texttt{TRUE} by default. For example:
\begin{verbatim}
> sc_a <- scalogram(signal = signal, dt = h,
+                   energy_density = FALSE)
\end{verbatim}
computes the (normalized) scalogram of \texttt{signal} given by \eqref{eq:sxsn} at a base $2$ power scales set constructed automatically and plots Figure~\ref{fig:sc} (a). The output \texttt{sc\_a} is a list with the following fields:
\begin{itemize}
\item \texttt{scalog} is a vector of length \texttt{length(scales)} with the values of the (normalized) scalogram at each scale.
\item \texttt{energy} is the total energy of \texttt{scalog} (i.e. its $L^2$-norm) if parameter \texttt{energy\_density} is \texttt{TRUE}.
\item \texttt{scales} and \texttt{fourier\_factor} are analogous to those in the output of function \texttt{cwt\_wst}.
\end{itemize}

If parameter \texttt{energy\_density} is set to \texttt{TRUE} (default value), then the scalogram is divided by the square root of the scales, converting it into an energy density measure (see Remark \ref{rmk:cwt}). For example, if we write:
\begin{verbatim}
> sc_b <- scalogram(signal = signal, dt = h)
\end{verbatim}
it plots Figure~\ref{fig:sc} (b), and then \texttt{sc\_b\$scalog} is in fact \texttt{sc\_a\$scalog / sqrt(scales)}.

\subsection{Inner Scalogram}

Given a compactly supported wavelet $\psi $ and $f$ a finite length signal with time domain $I=\left[ a,b\right] $, the \textit{(normalized) inner scalogram} of $f$ at scale $s$ is defined as
\begin{equation}
\label{eq:sinnerfs}
\mathcal{S}^{\textrm{inner}}f\left( s\right) =\left( \frac{1}{d(s)-c(s)} \int _{c(s)}^{d(s)} | \mathcal{W}f\left( u,s\right) |^2 \textrm{d}u\right) ^{1/2},
\end{equation}
where $\left[ c(s),d(s)\right] $ is the maximal subinterval in $I$ for which the support of $\psi _{u,s}$ is included in $I$ for all $u\in \left[ c(s),d(s)\right] $.
Hence, according to \eqref{eq:sinnerfs}, the \textit{(normalized) inner scalogram} of $x$ at scale $s$ is given by
\[
\mathcal{S}^{\mathrm{inner}}x(s) = \left( \frac{1}{n_2(s)-n_1(s)+1}\sum _{i=n_1(s)}^{n_2(s)} | \mathcal{W}x_i(s) | ^2 \right) ^{1/2},
\]
where $\left\{ n_1(s),\ldots ,n_2(s)\right\} $ is the maximal subset of time indices for which the support of $\psi _{i,s}$ is included in $\left[ t_0,t_N\right] $ for all $i\in \left\{ n_1(s),\ldots ,n_2(s)\right\} $.

This concept of inner scalogram can be extended to wavelets that do not have compact support, considering the effective support (see Remark \ref{rem:coi}) instead of the support. But we have to take into account that in this case, some theoretical results exposed in \cite{ben10} may not hold.

We can compute the (normalized) inner scalograms of a time series $x$ at a given set of scales by means of function \texttt{scalogram} setting the parameter \texttt{border\_effects} equal to \texttt{"INNER"}. Since Morlet, Paul and DoG wavelets are not compactly supported, it is considered the effective support given by the wavelet radius $r_w$.

\subsection{Windowed Scalogram}

The \textit{(normalized) windowed scalogram} of $f$ centered at time $t$ with time radius $\tau >0$ at scale $s$ is defined as
\begin{equation}
\label{eq:wstaufts}
\mathcal{WS}_{\tau }f(t,s) = \left( \frac{1}{2\tau }\int _{t-\tau } ^{t+\tau } | \mathcal{W}f\left( u,s\right) | ^2 \, \textrm{d}u \right)^{1/2}.
\end{equation}
It was introduced in \cite{bol17} and it allows to determine the relative importance of the different scales around a given time point. According to \eqref{eq:wstaufts}, the \textit{(normalized) windowed scalogram} of $x$ with time index radius $\tau \in \mathbb{N}$ at scale $s$ is given by the sequence
\begin{equation}
\label{eq:wsc}
\mathcal{WS}_{\tau }x_n(s) = \left( \frac{1}{2\tau +1}\sum _{i=n-\tau}^{n+\tau} | \mathcal{W}x_i(s) | ^2 \right) ^{1/2},
\end{equation}
where $n=\tau ,\ldots ,N-\tau $. In the particular case of $\tau =0$, \eqref{eq:wsc} coincides with $| \mathcal{W}x_n(s) |$.

\begin{figure}[tbp]
\begin{center}
  \includegraphics[width=1\textwidth]{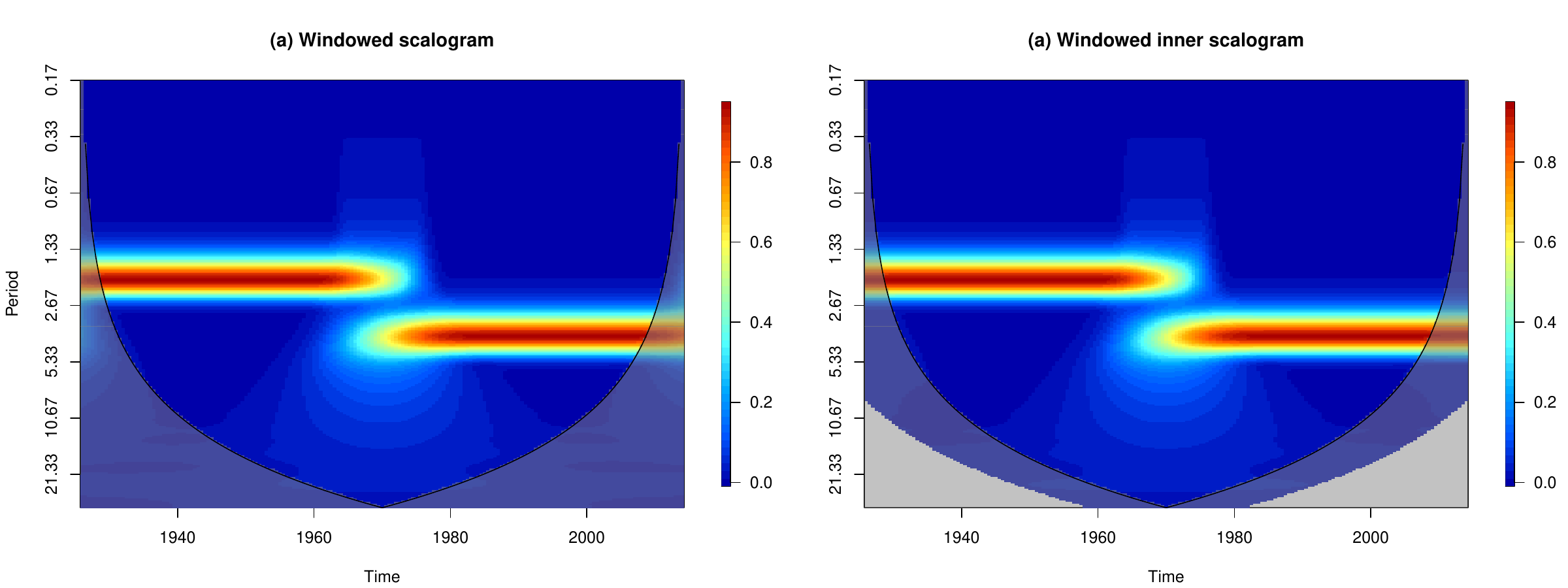}
\end{center}
\caption{Windowed scalogram (a) and windowed inner scalogram (b) of \texttt{signal}. The CoI is the shadowed region and, for the inner scalogram, the region where the scalogram cannot be computed is coloured in gray.}
\label{fig:wsc}
\end{figure}

We can compute the (normalized) windowed scalograms of a time series $x$ at a given set of scales by means of function \texttt{windowed\_scalogram}. Parameters \texttt{signal}, \texttt{dt}, \texttt{scales}, \texttt{powerscales}, \texttt{wname}, \texttt{wparam}, \texttt{waverad}, \texttt{border\_effects}, \texttt{energy\_density}, \texttt{makefigure}, \texttt{figureperiod}, \texttt{xlab}, \texttt{ylab} and \texttt{main} are analogous to those in function \texttt{scalogram}, and parameters \texttt{time\_values} and \texttt{zlim} are analogous to those in function \texttt{cwt\_wst}. For example:
\begin{verbatim}
> wsc <- windowed_scalogram(signal = signal, dt = h,
+                           windowrad = 72, delta_t = 6,
+                           time_values = c(time1, time2))
\end{verbatim}
computes the (normalized) windowed scalograms of \texttt{signal} with time index radius $\tau =\,$\texttt{windowrad} at a base $2$ power scales set constructed automatically. Moreover, it plots Figure~\ref{fig:wsc} (a). If \texttt{windowrad} is \texttt{NULL} (default value), then it is set to $\lceil (N+1)/20\rceil $. The parameter \texttt{delta\_t} is the index increment for the computation of the windowed scalograms, i.e. \eqref{eq:wsc} is computed only for $n$ from $\tau $ to $N-\tau $ by \texttt{delta\_t}. If \texttt{delta\_t} is \texttt{NULL} (default value) then it is taken $\lceil (N+1)/256\rceil $.

The output \texttt{wsc} is a list with the following fields:
\begin{itemize}
\item \texttt{tcentral} is the vector of times at which the windows are centered, i.e. the times of the form $t_n$ where $n$ goes from $\tau $ to $N-\tau $ by \texttt{delta\_t}.
\item \texttt{wsc} is a matrix of size \texttt{length(tcentral)}$\times$\texttt{length(scales)} containing the values of the windowed scalograms at each scale and at each central time.
\item \texttt{windowrad} is the time index radius $\tau $ used.
\item \texttt{scales}, \texttt{fourier\_factor} and \texttt{coi\_maxscale} are analogous to those in the output of function \texttt{cwt\_wst}.
\end{itemize}

\subsection{Windowed Inner Scalogram}

The \textit{(normalized) windowed inner scalogram} of $f$ centered at time $t$ with time radius $\tau >0$ at scale $s$ is defined as
\begin{equation}
\label{eq:wsinnerf}
\mathcal{WS}^{\mathrm{inner}}_{\tau }f(t,s) = \left( \frac{1}{d(t,s)-c(t,s)}\int _{c(t,s)} ^{d(t,s)} | \mathcal{W}f\left( u,s\right) | ^2 \, \textrm{d}u \right)^{1/2},
\end{equation}
where $\left[ c(t,s),d(t,s)\right] $ is the maximal subinterval in $\left[ t-\tau ,t+\tau \right] $ for which the effective support of $\psi _{u,s}$ is included in $I$ for all $u\in \left[ c(t,s),d(t,s)\right] $. Then, the \textit{(normalized) windowed inner scalogram} of $x$ with time index radius $\tau \in \mathbb{N}$ at scale $s$ is given by the sequence
\begin{equation}
\label{eq:wsinnerx}
\mathcal{WS}^{\mathrm{inner}}_{\tau }x_n(s) = \left( \frac{1}{n_2(n,s)-n_1(n,s)+1}\sum _{i=n_1(n,s)} ^{n_2(n,s)} | \mathcal{W}x_i(s) | ^2 \right)^{1/2},
\end{equation}
where $\left\{ n_1(n,s),\ldots ,n_2(n,s)\right\} $ is the maximal subset of time indices in $\left\{ n-\tau ,\ldots ,n+\tau \right\} $ for which the effective support of $\psi _{i,s}$ is included in $\left[ t_0,t_N\right] $ for all $i\in \left\{ n_1(n,s),\ldots ,n_2(n,s)\right\} $.

If \texttt{border\_effects} is set to \texttt{"INNER"} in function \texttt{windowed\_scalogram}, then the (normal\-ized) windowed inner scalograms are computed. For example:
\begin{verbatim}
> wsc <- windowed_scalogram(signal = signal, dt = h,
+                           windowrad = 72, delta_t = 6,
+                           border_effects = "INNER",
+                           time_values = c(time1, time2))
\end{verbatim}
computes the (normalized) windowed inner scalogram of \texttt{signal} with time index radius $\tau=\,$\texttt{windowrad}, at a base $2$ power scales set constructed automatically, and plots Figure~\ref{fig:wsc} (b). Note that in this figure, the ``CoI line'' is not a real CoI line, because if we consider inner scalograms, border effects are negligible. This line represents, at each time $t_n$, the maximum scale $s$ such that $n_1(n,s)=n-\tau $ and $n_2(n,s)=n+\tau $, and coincides with the CoI line of the (normalized) windowed scalogram.

\section{Windowed scalogram difference}
\label{sec4}

The \textit{windowed scalogram difference} (WSD) is a wavelet tool, introduced in \cite{bol17}, whose main objective is to compare time series by means of their respective windowed scalograms.

In order to consider differences between scalograms, it is convenient to use base $2$ power scales (\cite{bol17}) and hence, we must redefine them by making a change of variable. Thus, for example, from \eqref{eq:sxsn}, the (normalized) scalogram of a time series $x$ at \textit{log-scale} $k$ should be given by
\begin{equation}
\label{eq:sxkn}
\mathcal{S}x(k)= \left( \frac{1}{N+1}\sum _{i=0}^{N} | \mathcal{W}x_i(2^k) | ^2 \right) ^{1/2},
\end{equation}
where $k\in \mathbb{R}$ is the binary logarithm of the scale. From now on, $k$ will denote log-scales of scales $s$ in the sense that $2^k=s$.

The windowed scalogram difference of two signals
$f,g\in L^2\left( \mathbb{R}\right) $ centered at $(t,k)$ with time radius $\tau >0$ and log-scale
radius $\lambda >0$ is defined as
\begin{equation}
\label{eq:wsd}
\mathcal{WSD}_{\tau ,\lambda } fg(t,k)=\left( \int _{k-\lambda } ^{k+\lambda } \left( \frac{\mathcal{WS}_{\tau }f(t,\kappa ) - \mathcal{WS}_{\tau }g(t,\kappa )}{\mathcal{WS}_{\tau }f(t,\kappa )}\right) ^2 \, \mathrm{d}\kappa \right) ^{1/2}.
\end{equation}
The commutative version of \eqref{eq:wsd} is given by
\begin{equation}
\label{eq:wsdcomm}
\frac{1}{2}\left( \int _{k-\lambda } ^{k+\lambda } \left( \frac{\mathcal{WS}_{\tau }f(t,\kappa ) ^2- \mathcal{WS}_{\tau }g(t,\kappa )^2}{\mathcal{WS}_{\tau }f(t,\kappa )\, \mathcal{WS}_{\tau }g(t,\kappa )} \right)^2 \, \mathrm{d}\kappa \right) ^{1/2}.
\end{equation}
From \eqref{eq:wsd}, the \textit{windowed scalogram difference} (WSD) of two time series $x,y$ centered at log-scale $k$ with time index radius $\tau \in \mathbb{N}$ and log-scale radius $\lambda >0$ is given by the sequence
\begin{equation}
\label{eq:wsdxy}
\mathcal{WSD}_{\tau ,\lambda } xy_n(k)=\left( \int _{k-\lambda } ^{k+\lambda } \left( \frac{\mathcal{WS}_{\tau }x_n(\kappa ) - \mathcal{WS}_{\tau }y_n(\kappa )}{\mathcal{WS}_{\tau }x_n(\kappa )}\right) ^2 \, \mathrm{d}\kappa \right) ^{1/2},
\end{equation}
where $n=\tau ,\ldots ,N-\tau $. However, in practice, we work with a finite interval of log-scales that is discretized into $k_0,\ldots ,k_M$ with constant step. Thus, we can adapt \eqref{eq:wsdxy} to this situation so that it can be written as
\begin{equation}
\label{eq:wsdxy2}
\mathcal{WSD}_{\tau ,\lambda } xy_n(k_m)=
\left( \frac{2\lambda +1}{m_2-m_1+1} \sum _{i=m_1} ^{m_2} \left( \frac{\mathcal{WS}_{\tau }x_n(k_i) - \mathcal{WS}_{\tau }y_n(k_i)}{\mathcal{WS}_{\tau }x_n(k_i)}\right) ^2\right) ^{1/2},
\end{equation}
where $\lambda \in \mathbb{N}$ is the log-scale index radius, $m_1=\max \left\{ 0,m-\lambda \right\} $ and $m_2=\min \left\{ M,m+\lambda \right\} $. The factor $\frac{2\lambda +1}{m_2-m_1+1}$ is added to counteract the ``border effects'' in the log-scale interval that appear when $m-\lambda <0$ or $m+\lambda >M$, because in these cases, the number of addends is less than $2\lambda +1$. Moreover, according to \eqref{eq:wsdcomm}, a commutative version of \eqref{eq:wsdxy2} can be also considered.

We can compute the WSD \eqref{eq:wsdxy2} (or its commutative version) of two time series $x,y$ of the same length and time step by means of function \texttt{wsd}. Parameters \texttt{dt}, \texttt{windowrad}, \texttt{delta\_t}, \texttt{wname}, \texttt{wparam}, \texttt{waverad}, \texttt{border\_effects}, \texttt{energy\_density}, \texttt{makefigure}, \texttt{time\_values}, \texttt{figureperiod}, \texttt{xlab}, \texttt{ylab}, \texttt{main} and \texttt{zlim} are analogous to those in \texttt{windowed\_scalogram} function. For example:
\begin{verbatim}
> set.seed(12345) # For reproducibility
> N <- 1500
> time <- 0:N
> signal1 <- rnorm(n = N + 1, mean = 0, sd = 0.2) + sin(time / 10)
> signal2 <- rnorm(n = N + 1, mean = 0, sd = 0.2) + sin(time / 10)
> signal2[500:1000] = signal2[500:1000] + sin((500:1000) / 2)
> wsd <- wsd(signal1 = signal1, signal2 = signal2,
+            windowrad = 75, rdist = 14)
\end{verbatim}
computes the commutative WSD of \texttt{signal1} and \texttt{signal2} centered at $\left\{ k_0,\ldots ,k_M\right\} $, a log-scales set constructed automatically, with time index radius $\tau =\,$\texttt{windowrad} and log-scale index radius $\lambda =\,$\texttt{rdist}. If \texttt{windowrad} is \texttt{NULL} (default value) then it is set to $\lceil (N+1)/20\rceil $, and if \texttt{rdist} is \texttt{NULL} (default value) then it is set to $\lceil (M+1)/20\rceil $. The log-scales set can be defined by parameter \texttt{scaleparam}, that must be a vector of three elements with the minimum scale, the maximum scale and the number of suboctaves per octave. Moreover, logical parameter \texttt{commutative}, whose default value is \texttt{TRUE}, determines if it is computed the commutative version of the WSD.

\begin{remark}[Normalization]
\label{rem:norm}
The WSD compares the patterns of the windowed scalograms of two time series determining if they give similar weights (or energy) to the same scales. Another tool for comparing two time series is the squared wavelet coherence (\cite{tor98,tor99}), that measures the local linear correlation between them. So, these tools focus on different aspects: while the squared wavelet coherence does not take into account the magnitudes in the signals, for the WSD they are crucial. In fact, the WSD has sense only when the two time series considered are expressed in the same unit of measure or they are dimensionless. Otherwise, it will be necessary to somehow normalize the signals, but depending on the normalization method, some artifices could appear. For example, we can normalize the signals so that their scalograms have the same energy and, in this way, we can compare the relative contributions of each scale to the total energy. Another option is to normalize the signals so that their scalograms attain the same maximum value. Finally, it could be also useful to normalize the signals so that their scalograms reach the same value at a given reference scale.
\end{remark}

The normalization method can be chosen through parameter \texttt{normalize} in function \texttt{wsd}. It can be set to \texttt{"NO"} (default value), \texttt{"ENERGY"}, \texttt{"MAX"} or \texttt{"SCALE"}, according to each normalization method exposed in Remark \ref{rem:norm}. In this last option, the reference scale must be given by parameter \texttt{refscale}.

\begin{figure}[tbp]
\begin{center}
  \includegraphics[width=0.9\textwidth]{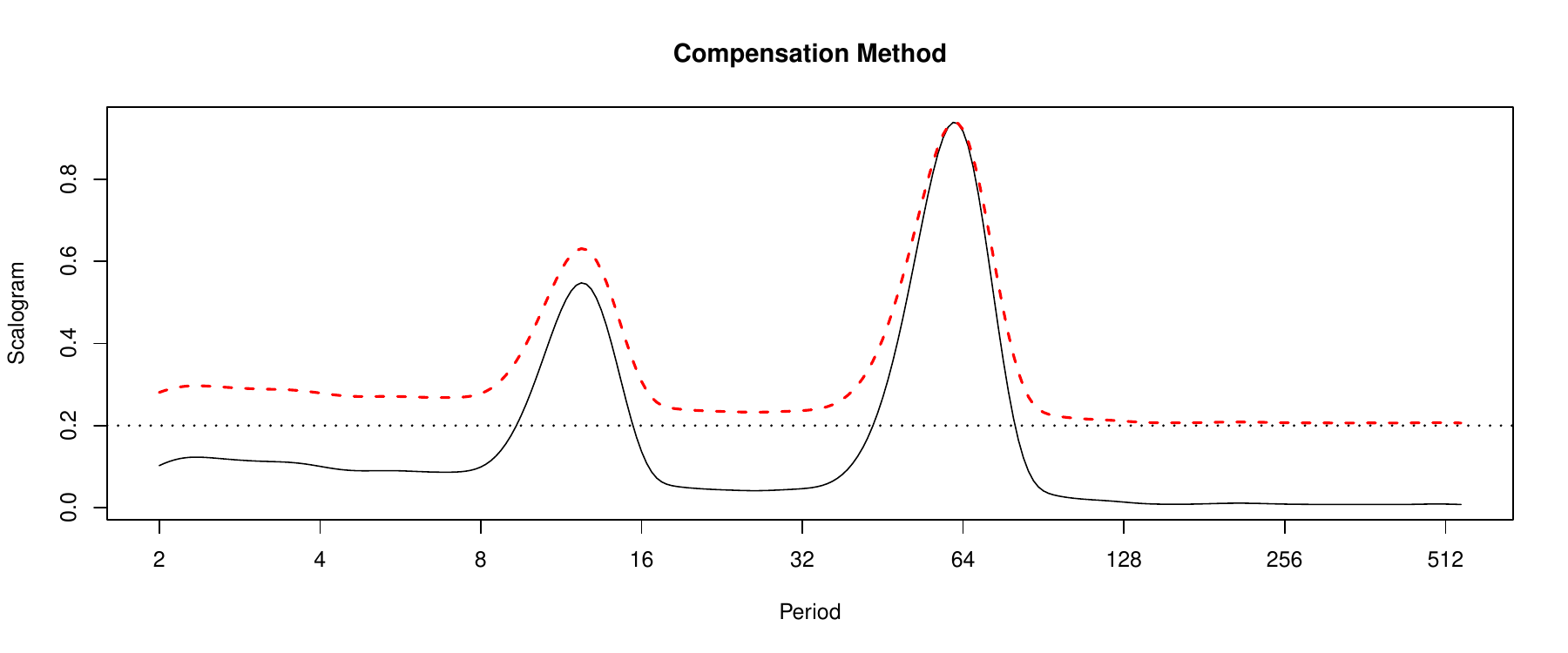}
\end{center}
\caption{Illustration of the compensation method exposed in Remark \ref{rem:com} to deal with near zero scalogram values in the computation of the WSD. Original scalogram of \texttt{signal2} (solid black line) is transformed into the compensated scalogram (dashed red line) for a compensation parameter $C=0.2$.}
\label{fig:comp}
\end{figure}

\begin{remark}[Near zero scalogram values]
\label{rem:com}
Some problems can arise in the WSD when a scalogram is zero or close to zero for a given log-scale because we are computing relative differences and hence, the WSD can take extremely high values or produce numerical errors. If we consider absolute differences this would not happen but, on the other hand, it would not be appropriate for scalogram values not close to zero.
A solution is to establish a threshold for the scalogram values above which a relative difference is computed, and below which a difference proportional to the absolute difference is computed (the proportionality factor would be determined by requiring continuity). This threshold can be interpreted as the relative amplitude of the noise in the scalograms.

Another solution is to substitute the original windowed scalograms $\mathcal{WS}_{\tau }x_n(s)$, $\mathcal{WS}_{\tau }y_n(s)$ by 
\begin{equation}
\label{eq:comp}
C+\left( 1-\frac{C}{\max}\right) \mathcal{WS}_{\tau }x_n(s), \qquad C+\left( 1-\frac{C}{\max}\right) \mathcal{WS}_{\tau }y_n(s),
\end{equation}
where
\[
\max =\max _{n,s}\left\{ \mathcal{WS}_{\tau }x_n(s),\mathcal{WS}_{\tau }y_n(s)\right\} ,
\]
and $C\geq 0$ is a relatively small value, called \textit{compensation} (see Figure~\ref{fig:comp}).
\end{remark}

Parameters \texttt{wscnoise} and \texttt{compensation} of function \texttt{wsd} allow us to deal with the near zero scalogram problem mentioned in Remark \ref{rem:com}. The first one is a value in $[0,1]$ and establishes the threshold from which a relative difference is computed. As particular cases, if it is $0$ then relative differences are always done, and if it is $1$ then absolute differences are always done. The default value is set to $0.02$. The second one determines the compensation $C$ of \eqref{eq:comp}, which is set to $0$ by default.

\begin{figure}[tbp]
\begin{center}
  \includegraphics[width=0.65\textwidth]{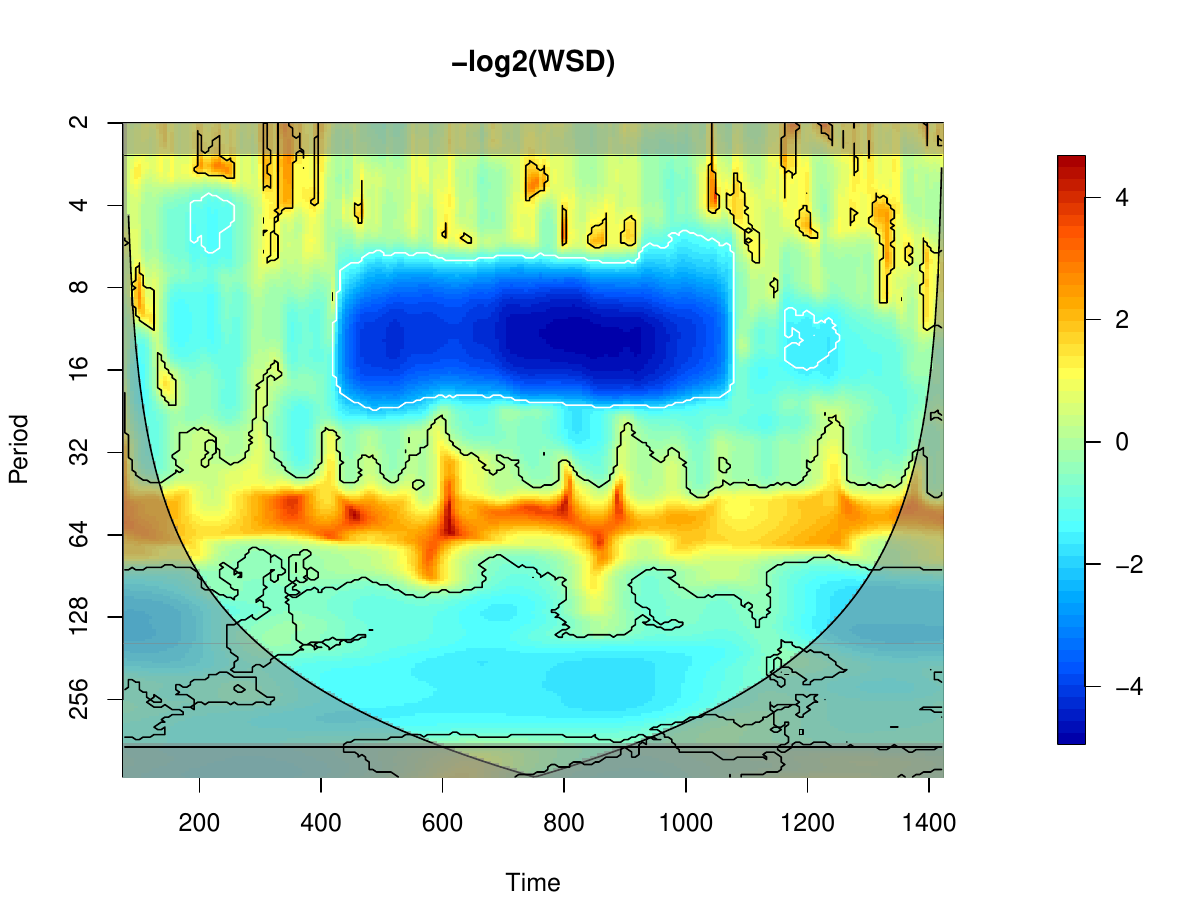}
\end{center}
\caption{Base $2$ logarithm of the inverse of the commutative WSD of \texttt{signal1} and \texttt{signal2} centered at a log-scales set constructed automatically, with time index radius $\tau =75$ and log-scale index radius $\lambda =14$. The significant contours are plotted in black (significantly high) and white (significantly low) lines, using 100 Monte Carlo simulations. Both time series are the same sinusoidal signal of period $20\pi$ plus different white noises with the same amplitude. Moreover, \texttt{signal2} has been manually modified for $500\leq t\leq 1000$, with the addition of another pure sin signal of period $4\pi$.  The red band around period $20\pi$ corresponds to the period both signals have in common while the dark blue region around period $4\pi$ corresponds to the period both differ.}
\label{fig:wsd}
\end{figure}

In practical situations, signals will be usually affected by random noises. Therefore it is necessary to determine whether the results obtained with the WSD are statistically significant or not.
In this package, we perform Monte Carlo simulations of the WSD (with the same parameter values) of random signals following a normal distribution with the same mean and standard deviation as the original ones. Then, we find the 95\% and 5\% quantiles to determine significantly high and low values respectively.
The number of Monte Carlo simulations is set by parameter \texttt{mc\_nrand} in function \texttt{wsd}, whose default value is $0$ (no significant contours are computed). For example:
\begin{verbatim}
> wsd <- wsd(signal1 = signal1, signal2 = signal2,
+            mc_nrand = 100, parallel = TRUE)
\end{verbatim}
computes the same WSD as before, but determines which values are significant using $100$ Monte Carlo simulations, plotting Figure~\ref{fig:wsd}. Parameter \texttt{parallel} enables parallel computations improving considerably the execution time for high values of \texttt{mc\_nrand}.

Finally, the output of \texttt{wsd} is a list with the following fields:
\begin{itemize}
\item \texttt{wsd} is a matrix of size \texttt{length(tcentral)}$\times $\texttt{length(scales)} containing the values of the WSD at each scale and at each central time.
\item \texttt{rdist} is the log-scale index radius $\lambda $ used.
\item \texttt{signif95} and \texttt{signif05} are logical \texttt{length(tcentral)}$\times $\texttt{length(scales)} matrices, that determine if the corresponding values of the \texttt{wsd} matrix are significantly high or low respectively, following the 95\% and 5\% quantiles method described above.
\item \texttt{tcentral}, \texttt{scales}, \texttt{windowrad}, \texttt{fourier\_factor} and \texttt{coi\_maxscale} are analogous to those in the output of function \texttt{windowed\_scalogram}.
\end{itemize}

With respect to the output image, it is plotted the base $2$ logarithm of the inverse of the WSD because in this way high values represent small differences (i.e. high similarity) and low values represent large differences (i.e. low similarity) (\cite{bol17}).

\section{Scale index and windowed scale index}
\label{sec5}

Periodicity is one of the most basic characteristics to be determined in a time series study. Mathematically, the definition is clear: a time series $f$ is periodic of period $T$ whenever $f(t+T) = f(t)$ for all $t$, and a time series that fails to be periodic is a non-periodic signal. However, within this definition, there are very different types of non-periodic signals (e.g. stochastic, quasi-periodic, chaotic signals), and an interesting question to analyze is how much non-periodic a time series is. Within this regard, the \emph{scale index} and the \emph{windowed scale index} (\cite{ben10,bol20}) are two wavelet tools that give a satisfactory answer to this question.

The \textit{scale index} of a signal $f\in L^2\left( \mathbb{R}\right) $ in the scale interval $\left[ s_0,s_1\right] $ is defined as the quotient
\begin{equation}
\label{eq:sif}
i_{\textrm{scale}}f=\frac{\mathcal{S}f(s_{\textrm{min}})}{\mathcal{S}f(s_{\textrm{max}})},
\end{equation}
where $s_{\textrm{max}}\in \left[ s_0,s_1\right] $ is the smallest scale such that $\mathcal{S}f(s)\leq \mathcal{S}f(s_{\textrm{max}})$ for all $s\in \left[ s_0,s_1\right] $, and $s_{\textrm{min}}\in \left[ s_{\textrm{max}},2s_1\right] $ is the smallest scale such that $\mathcal{S}f(s_{\textrm{min}})\leq \mathcal{S}f(s)$ for all $s\in \left[ s _{\textrm{max}}, 2s_1 \right] $ (\cite{ben10,bol20}). Hence, according to \eqref{eq:sif}, the \textit{scale index} of a time series $x$ in the scale interval $\left[ s_0,s_1\right] $ is given by
\begin{equation}
\label{eq:six}
i_{\textrm{scale}}x=\frac{\mathcal{S}x(s_{\textrm{min}})}{\mathcal{S}x(s_{\textrm{max}})},
\end{equation}
where $s_{\textrm{max}}$ and $s_{\textrm{min}}$ are defined analogously.

The scale index is a quantity in $[0,1]$ and measures the degree of non-periodicity of a signal in a given scale interval $\left[ s_0,s_1\right] $: it is close to zero for periodic and quasi-periodic signals, and close to one for highly non-periodic signals. The choice of the scale interval $\left[ s_0,s_1\right] $ is very important, and it should contain all the relevant scales that we want to study.

\begin{remark}[No energy density]
\label{rmk:noed}
The correction exposed in Remark \ref{rmk:cwt} should not be carried out because the scalogram of a white noise signal is more or less constant at all scales giving a scale index close to $1$ for any scale interval $\left[ s_0,s_1\right] $, and this is the property that we want to preserve. If, on the other hand, we apply the correction for converting the scalogram into an ``energy density'' measure, the scale index of a white noise signal would tend to zero as we increase $s_1$ and this is not desirable (\cite{bol20}).
\end{remark}

We can compute the scale index of a time series $x$ by means of function \texttt{scale\_index}. Parameters \texttt{signal}, \texttt{dt}, \texttt{scales}, \texttt{powerscales}, \texttt{wname}, \texttt{wparam}, \texttt{waverad}, \texttt{border\_effects}, \texttt{makefigure}, \texttt{figureperiod}, \texttt{xlab}, \texttt{ylab} and \texttt{main} are analogous to those in \texttt{scalogram} function. Note that, according to Remark \ref{rmk:noed}, there is no parameter \texttt{energy\_density} because scalograms must be computed without this correction. For example:
\begin{verbatim}
> set.seed(12345) # For reproducibility
> N <- 999
> h <- 1 / 8
> time <- seq(from = 0, to = N * h, by = h)
> signal_si <- sin(pi * time) + rnorm(n = N + 1, mean = 0, sd = 2)
> s0 <- 1
> s1 <- 4
> si <- scale_index(signal = signal_si, dt = h,
+                   scales = c(s0, 2 * s1, 24), s1 = s1,
+                   border_effects = "INNER", makefigure = FALSE)
\end{verbatim}
computes the scale index of \texttt{signal\_si} in the scale interval $\left[ s_0,s_1\right] $ where $s_0=1$ and $s_1=4$. The parameter \texttt{scales} determines the scales set at which the scalograms are computed: in this case, it is a base $2$ power scales set from $s_a=s_0$ to $s_b=2s_1$ with $24$ suboctaves per octave. Note that we take $s_b=2s_1$ according to the definition of the scale index, because $s_b$ can not be lesser than $2s_1$ and there is no need for $s_b$ to be greater than $2s_1$. Moreover, function \texttt{scale\_index} takes $s_0$ equal to the lowest scale $s_a$ always. If \texttt{scales = NULL} (default value), then the scalograms are computed at an automatically constructed set of scales with $s_a$ equal to the scale whose equivalent Fourier period is $2h$, and $s_b=2s_1$.

\begin{figure}[tbp]
\begin{center}
  \includegraphics[width=0.8\textwidth]{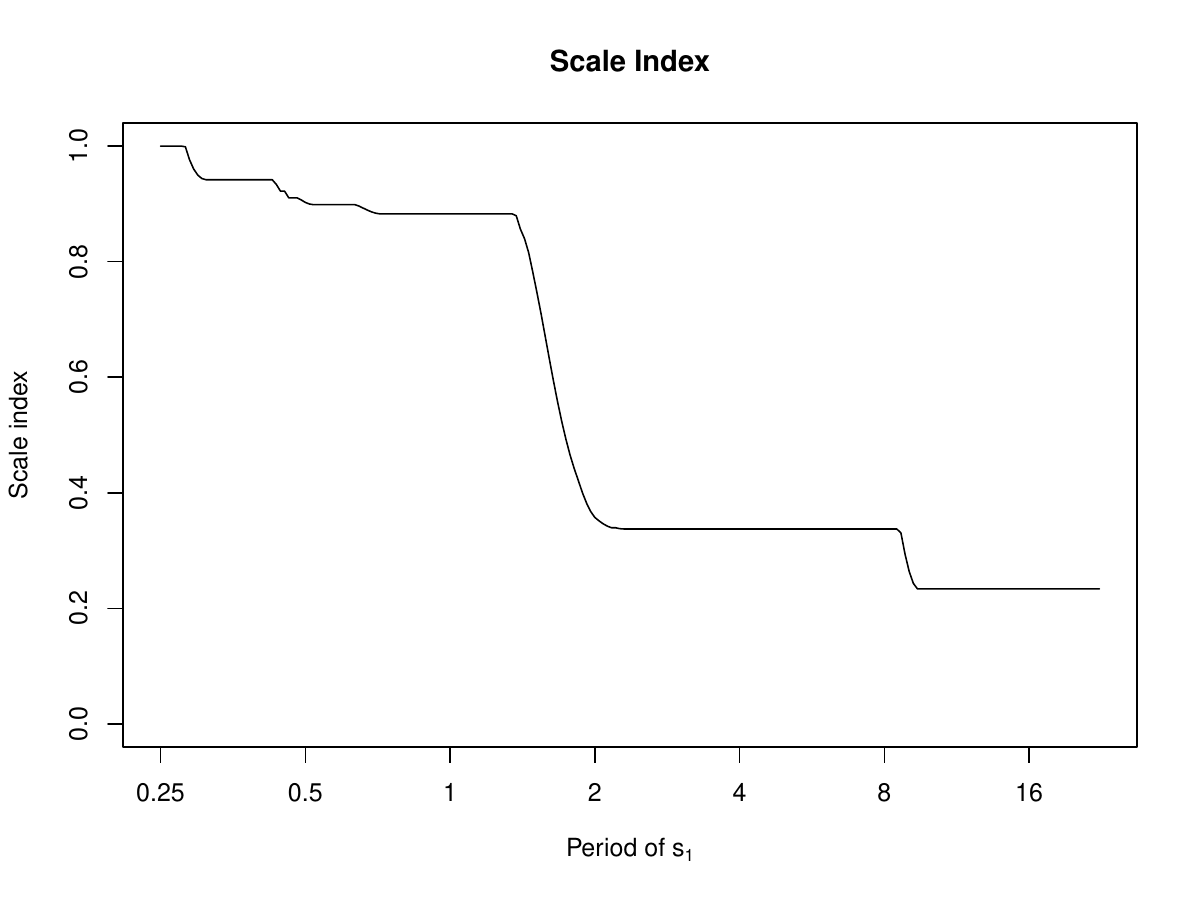}
\end{center}
\caption{Scale indices of \texttt{signal\_si} in scale intervals $\left[ s_0,s_1\right] $ where $s_0$ is the scale whose equivalent Fourier period is $0.25$ and $s_1$ varies. The signal is a pure $\sin $ of period $2$ plus a noise term. Thus, for values of $s_1$ lower than $2$, the scale index is very high because the scalogram still has not considered this period. At $s_1=2$, there is a sudden drop in the scale index and from that point onwards, as $s_1$ increases, the scale index decreases until it reaches a stable plateau (at $0.2$ approximately) for large values of $s_1$.}
\label{fig:sidx}
\end{figure}

We can also compute the scale indices of a signal in scale intervals $\left[ s_0,s_1\right] $ for different values of $s_1$ assigning a vector of scales to the parameter \texttt{s1}. Thus,
\begin{verbatim}
> maxs1 <- 4
> si <- scale_index(signal = signal_si, dt = h,
+                   scales = c(s0, 2 * maxs1, 24),
+                   s1 = pow2scales(c(s0, maxs1, 24)),
+                   border_effects = "INNER")
\end{verbatim}
computes the scale indices of \texttt{signal\_si} in scale intervals $\left[ s_0,s_1\right] $ where $s_0=1$ and $s_1$ varies in a base $2$ power scales set from $1$ to $4$ with 24 suboctaves per octave. Moreover, if \texttt{s1 = NULL} (default value), then \texttt{s1} is automatically computed as a base $2$ power scales set from $s_0$ to $s_b/2$. If \texttt{scales} is also \texttt{NULL}, then $s_b=Nh/2r_w$ as usual, where $r_w$ is the corresponding wavelet radius (see Remark \ref{rem:coi}). Hence,
\begin{verbatim}
> si <- scale_index(signal = signal_si, dt = h, border_effects = "INNER")
\end{verbatim}
computes the scale indices of \texttt{signal\_si} in scale intervals $\left[ s_0,s_1\right] $ where $s_0$ is the scale whose equivalent Fourier period is $2h$ and $s_1$ varies in a base $2$ power scales set from $s_0$ to $Nh/r_w$. Moreover, it returns a plot like Figure~\ref{fig:sidx}. It is important to remark that if \texttt{s1} are not base 2 power scales then \texttt{powerscales} must be \texttt{FALSE}. For example,
\begin{verbatim}
> si <- scale_index(signal = signal_si, dt = h,
+                   s1 = seq(from = s0, to = maxs1, by = 0.1),
+                   powerscales = FALSE, border_effects = "INNER")
\end{verbatim}
computes the scale indices of \texttt{signal\_si} in scale intervals $\left[ s_0,s_1\right] $ where $s_0=1$ and $s_1$ varies linearly from $1$ to $4$ with step $0.1$. In this case, since \texttt{scales} are not given, they are constructed automatically in a linear form, since \texttt{powerscales} must be \texttt{FALSE}.

\begin{figure}[tbp]
\begin{center}
  \includegraphics[width=\textwidth]{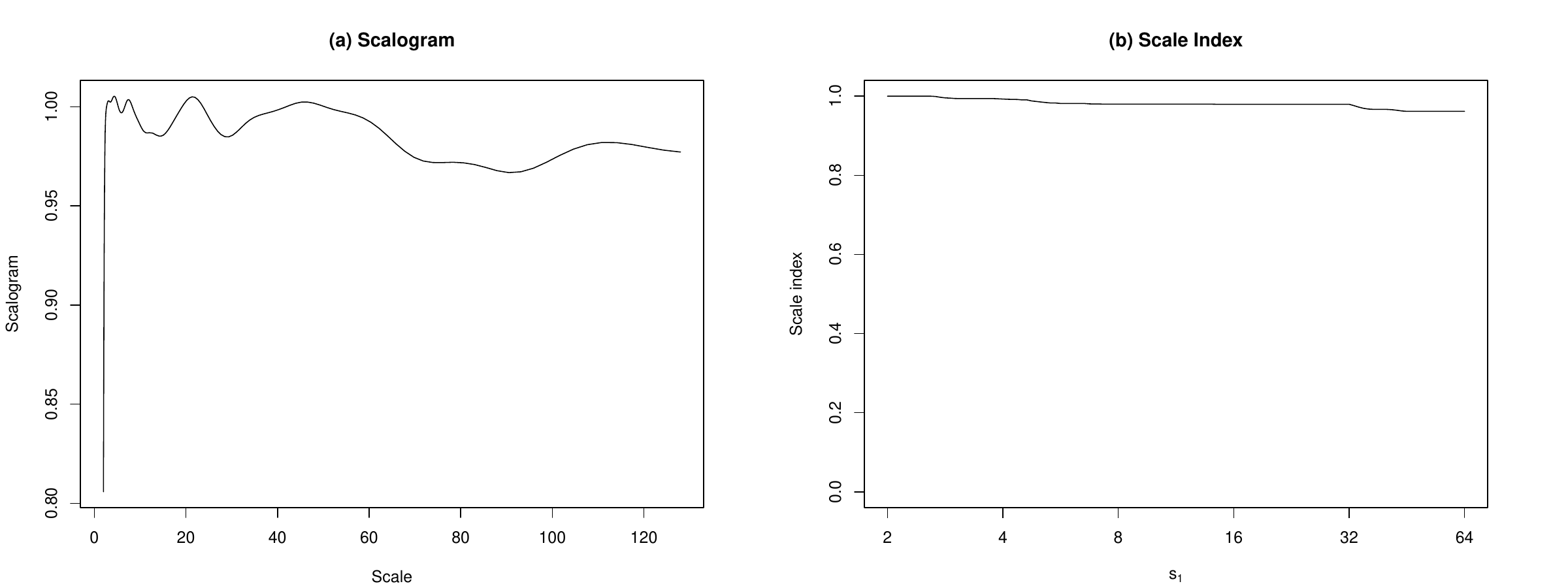}
\end{center}
\caption{(a) Average of 100 scalograms of noise signals. (b) Scale indices computed from this averaged scalogram. The scale indices are very close to $1$, indicating a high degree of non-periodicity.}
\label{fig:si_prng}
\end{figure}

Alternatively, we can compute the scale indices directly from a scalogram instead of giving the original signal by means of parameter \texttt{scalog}. In this case, we must give the scales at which the scalogram has been computed. Thus, we can compute the scale indices of an artificially constructed scalogram which does not necessarily have to correspond to any signal. For example, we can compute the scale indices corresponding to the average of $100$ scalograms of noise signals:
\begin{verbatim}
> set.seed(12345) # For reproducibility
> N <- 1000
> nrand <- 100
> X <- matrix(rnorm(N * nrand), nrow = N, ncol = nrand)
> scales = pow2scales(c(2, 128, 24))
> ns = length(scales)
> sc_list <- apply(X, 2, scalogram, scales = scales,
+                  border_effects = "INNER",
+                  energy_density = FALSE, makefigure = FALSE)
> sc_matrix <- matrix(unlist(lapply(sc_list, "[[", "scalog")),
+                     nrow = ns, ncol = nrand)
> sc_mean <- apply(sc_matrix, 1, mean)
> s1 = pow2scales(c(2, 64, 24))
> si_mean <- scale_index(scalog = sc_mean, scales = scales, s1 = s1,
+                        figureperiod = FALSE, plot_scalog = TRUE)
\end{verbatim}
This code also returns figures like those in Figure \ref{fig:si_prng}. The logical parameter \texttt{plot\_scalog} is used for plotting the scalogram from which the scale indices are computed.

The output of \texttt{scale\_index} is a list with the following fields:
\begin{itemize}
\item \texttt{si} is a vector with the scale indices, for each value of $s_1$.
\item \texttt{s0} is the scale $s_0$.
\item \texttt{s1} is a vector with the scales $s_1$.
\item \texttt{smax} and \texttt{smin} are vectors with the scales $s_{\textrm{max}}$ and $s_{\textrm{min}}$ respectively, for each value of $s_1$.
\item \texttt{scalog} is the the scalogram  $\mathcal{S}x$ from which the scale indices are computed.
\item \texttt{scalog\_smax} and \texttt{scalog\_smin} are vectors with the scalogram values $\mathcal{S}x(s_{\textrm{max}})$ and $\mathcal{S}x(s_{\textrm{min}})$ respectively, for each value of $s_1$.
\item \texttt{fourierfactor} is the scalar used to transform scales into Fourier periods (see Remark \ref{rem:ff}). 
\end{itemize}

\subsection{Windowed Scale Index}

As was mentioned in the introduction, wavelet analysis is a very useful tool for non-stationary time series. If we are interested in analyzing the non-periodicity of a non-stationary time series, we should be aware that the scale index is going to give us a single number between 0 and 1 which represents the degree of non-periodicity of the signal in the overall time interval of interest. However we may be interested in how this degree of non-periodicity is changing along this interval. To this aim, \cite{bol20} introduced the \emph{windowed scale index}, which uses the windowed scalogram in order to obtain scale indices for different time and scale intervals. 

In particular, the \textit{windowed scale index} of $f$ in the scale interval $[s_0,s_1]$ centered at time $t$ with time radius $\tau >0$ is defined as
\begin{equation}
\label{eq:wsif}
wi_{\textrm{scale},\tau }f(t) = \frac{\mathcal{WS}_{\tau }f\left( t,s_{\textrm{min}}\right) }{\mathcal{WS}_{\tau }f\left( t,s_{\textrm{max}}\right) },
\end{equation}
where, analogously to \eqref{eq:sif}, $s_{\textrm{max}}$ is the smallest scale such that $\mathcal{WS}_{\tau }f(t,s)\leq \mathcal{WS}_{\tau }f\left( t,s_{\textrm{max}}\right) $ for all $s\in [s_0,s_1]$, and $s_{\textrm{min}}$ is the smallest scale such that $\mathcal{WS} _{\tau }f\left( t,s _{\textrm{min}}\right) \leq \mathcal{WS} _{\tau}f(t,s)$ for all $s \in \left[ s _{\textrm{max}}, 2s_1 \right] $ (\cite{bol20}). Finally, according to \eqref{eq:wsif}, the \textit{windowed scale index} of a time series $x$ in the scale interval $[s_0,s_1]$ with time index radius $\tau \in \mathbb{N}$ is given by the sequence
\[
wi_{\textrm{scale},\tau }x_n = \frac{\mathcal{WS}_{\tau }x_n\left(s_{\textrm{min}}\right) }{\mathcal{WS}_{\tau }x_n\left(s_{\textrm{max}}\right) },
\]
where $n=\tau ,\ldots ,N-\tau $, and $s_{\textrm{max}},s_{\textrm{min}}$ are defined analogously to \eqref{eq:wsif}.

\begin{remark}[Inner scalograms]
Although in the computation of the scale index it is recommended the use of (normalized) inner scalograms in order to fulfil some theoretical results and avoid border effects, this recommendation is less important in the case of the windowed scale index, because for long time series and relatively small time radii there would be no relevant border effects in most of the windowed scalograms.
\end{remark}

\begin{figure}[tbp]
\begin{center}
  \includegraphics[width=0.8\textwidth]{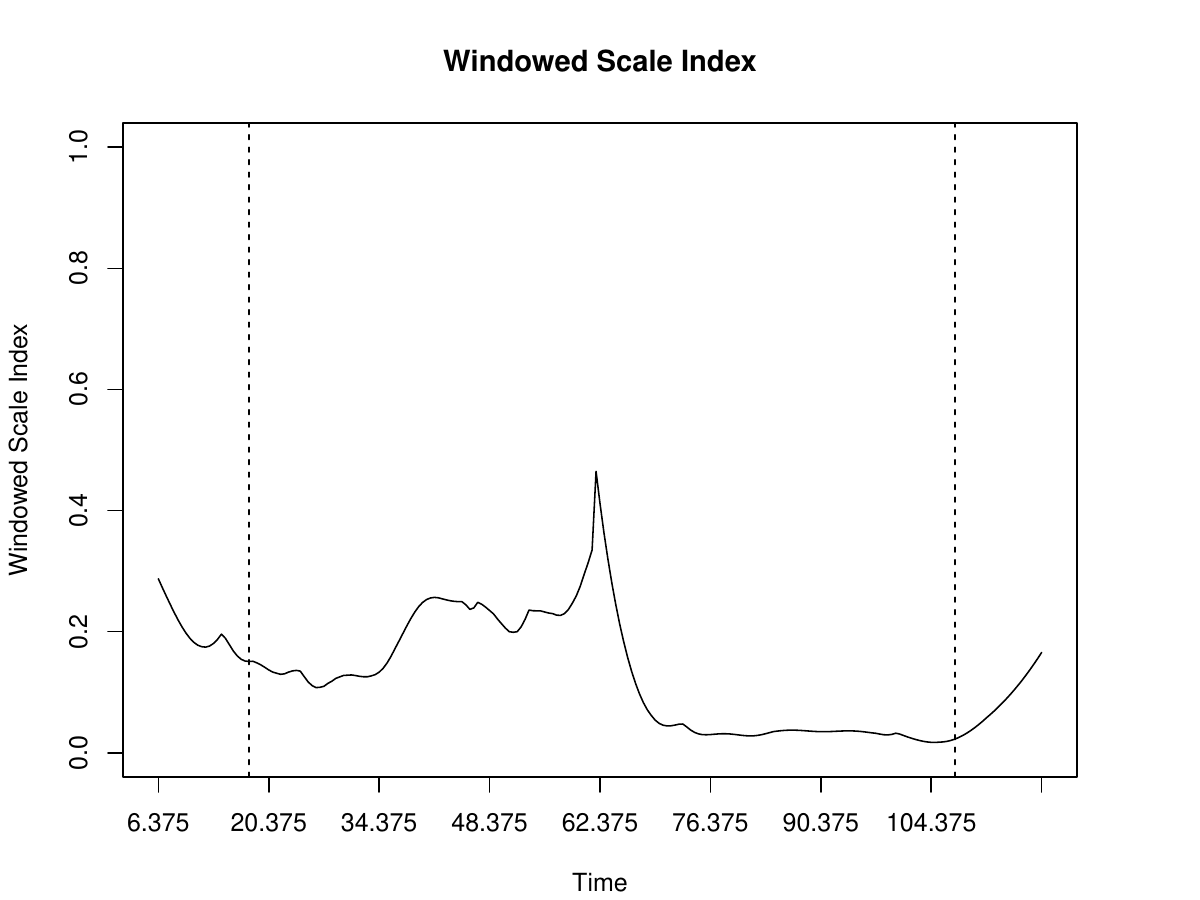}
\end{center}
\caption{Windowed scale index of \texttt{signal\_wsi} in a scale interval $\left[ 1,4\right] $ and with time index radius $\tau =50$. The dashed vertical lines represent the CoI limits. This time series is the concatenation of two sinusoidal signals of periods $2$ and $4$, modified with two white noises of different variance. In the first part, where the noise has a higher standard deviation, the windowed scale index is also higher. Moreover, it can be seen how the windowed scale index captures the moment of change in the noise.}
\label{fig:wsidx4}
\end{figure}

By means of function \texttt{windowed\_scale\_index}, we can compute the windowed scale index of a time series $x$. As usual, parameters \texttt{signal}, \texttt{dt}, \texttt{scales}, \texttt{powerscales}, \texttt{windowrad}, \texttt{delta\_t}, \texttt{wname}, \texttt{wparam}, \texttt{waverad}, \texttt{border\_effects}, \texttt{makefigure}, \texttt{time\_values}, \texttt{figure\-period}, \texttt{xlab}, \texttt{ylab}, \texttt{main} and \texttt{zlim} are analogous to those in function \texttt{windowed\_scalogram}. Moreover, parameter \texttt{s1} is analogous to that in function \texttt{scale\_index}. For example:
\begin{verbatim}
> set.seed(12345) # For reproducibility
> s0 <- 1
> s1 <- 4
> signal1_wsi <- sin(pi * time[1:500]) +
+                rnorm(n = 500, mean = 0, sd = 2)
> signal2_wsi <- sin(pi * time[501:1000] / 2) +
+                rnorm(n = 500, mean = 0, sd = 0.5)
> signal_wsi <- c(signal1_wsi, signal2_wsi)
> wsi <- windowed_scale_index(signal = signal_wsi, dt = h,
+                             scales = c(s0, 2 * s1, 24), s1 = s1,
+                             windowrad = 50,
+                             time_values = time)
\end{verbatim}
computes the windowed scale index of \texttt{signal\_wsi} in a scale interval $\left[ s_0,s_1\right] $ where $s_0=1$ and $s_1=4$. The time index radius $\tau$ is given by the parameter \texttt{windowrad}. If it is \texttt{NULL} (default value), then it is set to $\lceil (N+1)/20\rceil $ that, in this case, coincides with the value of \texttt{windowrad}. Moreover, it returns a plot like Figure~\ref{fig:wsidx4}.

\begin{figure}[tbp]
\begin{center}
  \includegraphics[width=0.8\textwidth]{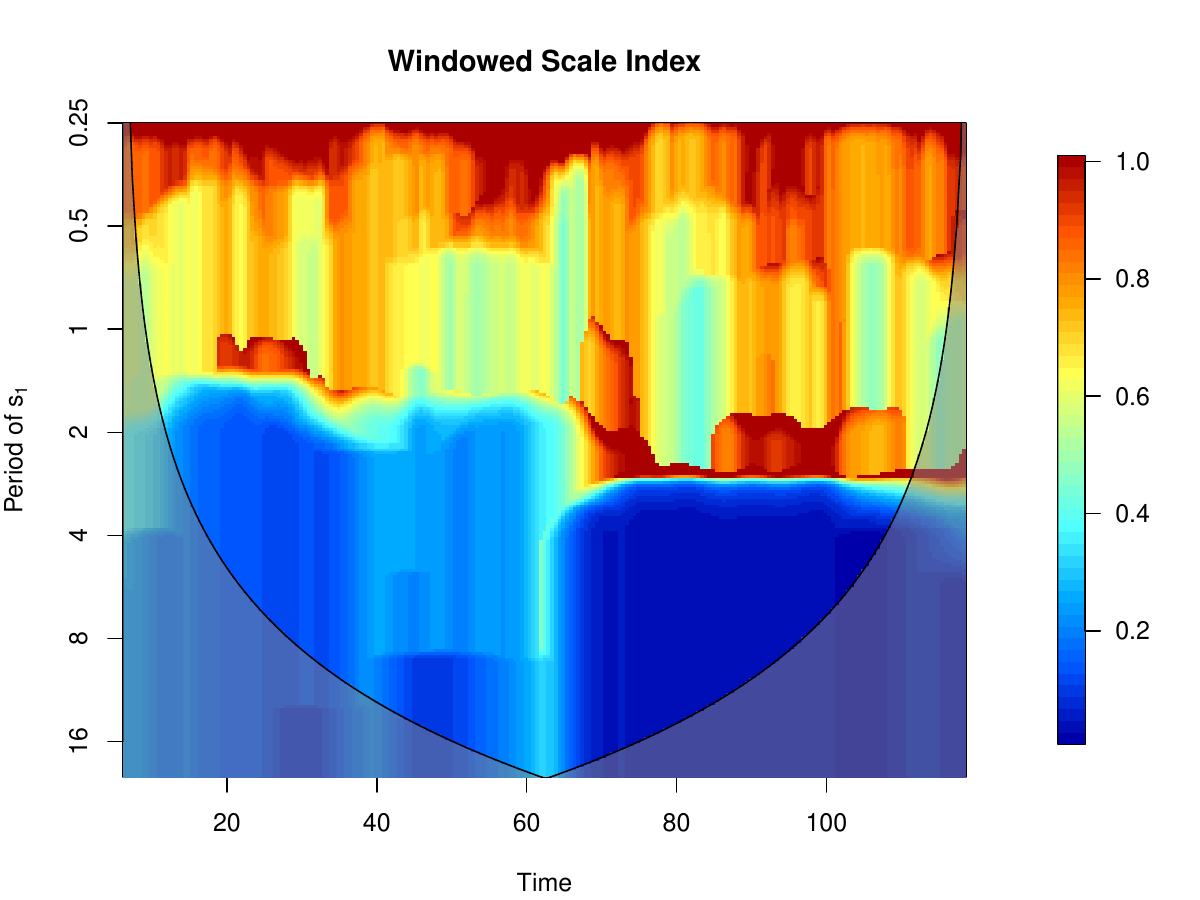}
\end{center}
\caption{Windowed scale indices of \texttt{signal\_wsi} in scale intervals $\left[ s_0,s_1\right] $ where $s_0$ is the scale whose equivalent Fourier period is $0.25$ and $s_1$ varies, with time index radius $\tau =50$. This plot also shows that $s_1$ should be at least $4$ for the scale indices to capture all relevant periods.}
\label{fig:wsidx}
\end{figure}

We can compute the windowed scale indices for different values of $s_1$ assigning a vector of scales to the parameter \texttt{s1}. It is important to remark that if \texttt{s1} are not base 2 power scales, then \texttt{powerscales} must be \texttt{FALSE}. If \texttt{s1 = NULL} and/or \texttt{scales = NULL} (default values), then they are automatically computed in the same way as it is done in function \texttt{scale\_index}. So,
\begin{verbatim}
> wsi <- windowed_scale_index(signal = signal_wsi, dt = h,
+                             time_values = time)
\end{verbatim}
computes the windowed scale indices of \texttt{signal\_wsi} in scale intervals $\left[ s_0,s_1\right] $ where $s_0$ is the scale whose equivalent Fourier period is $2h$ and $s_1$ varies in a base $2$ power scales set from $s_0$ to $Nh/r_w$. The time index radius $\tau $ is taken automatically as $\lceil (N+1)/20\rceil =50$. It also returns a plot, like Figure~\ref{fig:wsidx}.

\begin{figure}[tbp]
\begin{center}
  \includegraphics[width=\textwidth]{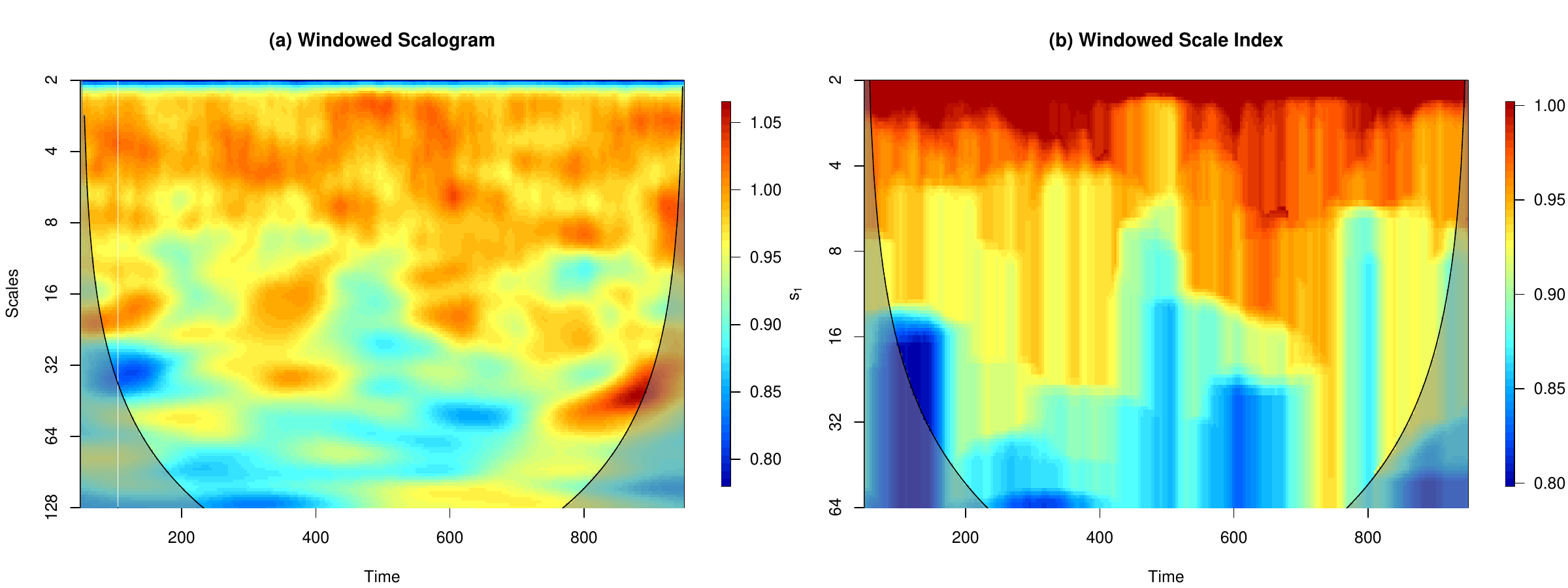}
\end{center}
\caption{(a) Average of 100 windowed scalograms of noise signals. (b) Windowed scale indices computed from this averaged windowed scalogram. The windowed scale indices are always close to $1$, indicating a high degree of non-periodicity.}
\label{fig:wsi_prng}
\end{figure}

Alternatively, we can compute the windowed scale indices directly from a windowed scalogram instead of giving the original signal by means of parameter \texttt{wsc}. This parameter must be equal to a matrix of size (number of central times)$\times$(number of scales), as it is returned by the \texttt{windowed\_scalogram} function. In this case, we must give the scales at which the windowed scalogram \texttt{wsc} has been computed and, in addition, we can give the cone of influence by means of parameter \texttt{wsc\_coi}, that must be a vector containing the values of the maximum scale at each central time from which there are border effects in \texttt{wsc}. Thus, we can compute the windowed scale indices of an artificially constructed windowed scalogram as it was shown in the case of the scale index. Taking the same example, we can compute the windowed scale indices corresponding to the average of $100$ windowed scalograms of noise signals:
\begin{verbatim}
> set.seed(12345) # For reproducibility
> N <- 1000
> nrand <- 100
> X <- matrix(rnorm(N * nrand), nrow = N, ncol = nrand)
> scales = pow2scales(c(2, 128, 24))
> ns = length(scales)
> wsc_list <- apply(X, 2, windowed_scalogram, scales = scales,
+                   energy_density = FALSE, makefigure = FALSE)
> tcentral <- wsc_list[[1]]$tcentral
> ntc <- length(tcentral)
> wsc_matrix <- array(unlist(lapply(wsc_list, "[[", "wsc")),
+                     c(ntc, ns, nrand))
> wsc_mean <- apply(wsc_matrix, 1:2, mean)
> wsc_coi <- wsc_list[[1]]$coi_maxscale
> wsi_mean <- windowed_scale_index(wsc = wsc_mean, wsc_coi = wsc_coi,
+                                  scales = scales, time_values = tcentral,
+                                  figureperiod = FALSE, plot_wsc = TRUE)
\end{verbatim}
This code also returns figures like those in Figure \ref{fig:wsi_prng}. The logical parameter \texttt{plot\_wsc} is used for plotting the windowed scalogram from which the windowed scale indices are computed.

The output of \texttt{windowed\_scale\_index} is a list with the following fields:
\begin{itemize}
\item \texttt{wsi} is a matrix of size \texttt{length(tcentral)}$\times $\texttt{length(s1)} with the windowed scale indices at each $s_1$ and at each central time.
\item \texttt{wsc} is a matrix of size \texttt{length(tcentral)}$\times $\texttt{length(scales)} with the windowed scalograms from which the windowed scale indices are computed. Note that scales greater than \texttt{2*max(s1)} are not necessary and they are internally removed from \texttt{scales}.
\item \texttt{s0}, \texttt{s1}, \texttt{smax}, \texttt{smin}, \texttt{scalog\_smax} and \texttt{scalog\_smin} are analogous to those in the output of function \texttt{scale\_index}.
\item \texttt{tcentral}, \texttt{windowrad}, \texttt{fourierfactor} and \texttt{coi\_maxscale} are analogous to those in the output of function \texttt{windowed\_scalogram}.
\end{itemize}

\section{Examples and applications}
\label{sec6} 

\subsection{Windowed scalogram difference and clustering}

As an application, we are going to show an example of how to define a dissimilarity measure from the windowed scalogram difference (WSD), which can then be applied to perform time series clustering. We are going to use the \texttt{interest.rates} time series from package \textbf{TSclust} (\cite{jss14}), which consists on $215$ observations of the monthly long-term interest rates (10-year bonds) from January 1995 to November 2012 of several countries.

\begin{figure}[tbp]
\begin{center}
  \includegraphics[width=\textwidth]{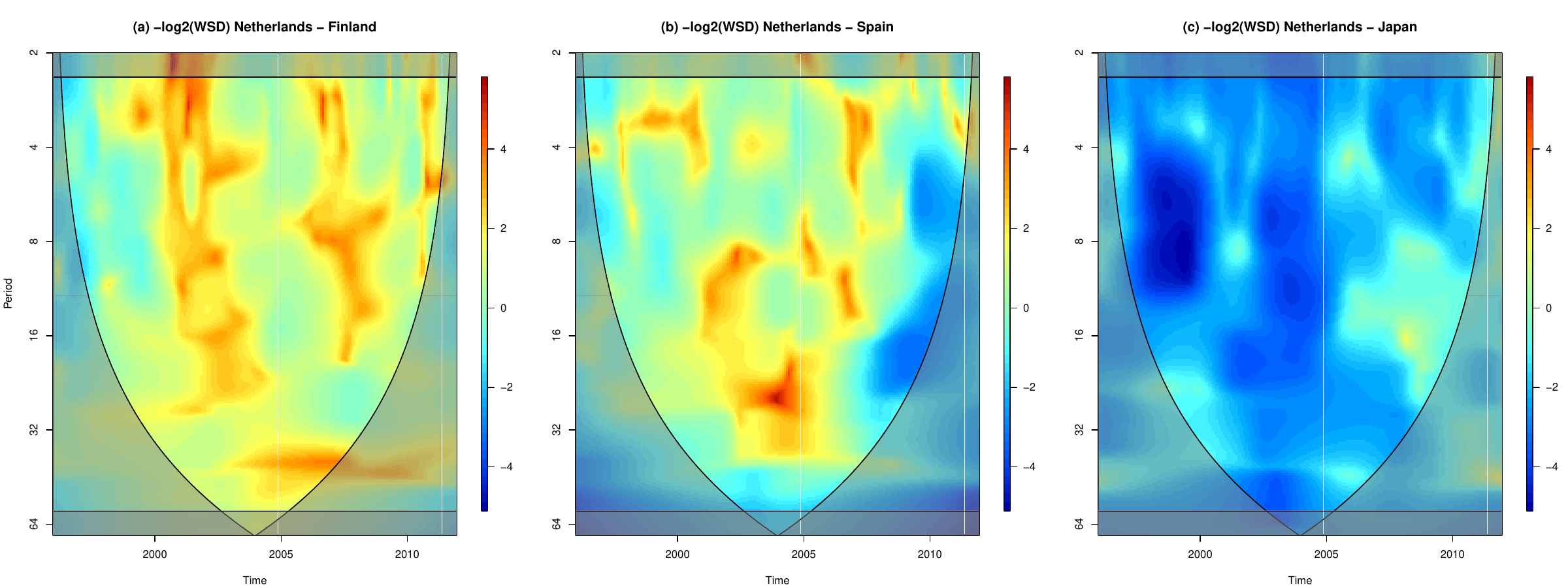}
\end{center}
\caption{Plots of base $2$ logarithms of the inverse of the commutative WSD of returns of Netherlands and (a) Finland, (b) Spain and (c) Japan. The corresponding dissimilarity measures of these pairs are $0.7395$, $1.6279$ and $2.819$ respectively. Red zones indicate time-scale regions where the two signals are more similar, while blue zones correspond to less similarity between the signals. Note that Netherlands and Finland are two countries whose economies are similar in both time and scale (plot (a)) but, on the other hand, Netherlands has a very different economic behaviour than Japan (plot (c)). The big blue spot in plot (b) corresponds to the 2008 financial crisis, which hit Spain harder than the Netherlands.}
\label{fig:interestwsd}
\end{figure}

\begin{figure}[tbp]
\begin{center}
  \includegraphics[width=.8\textwidth]{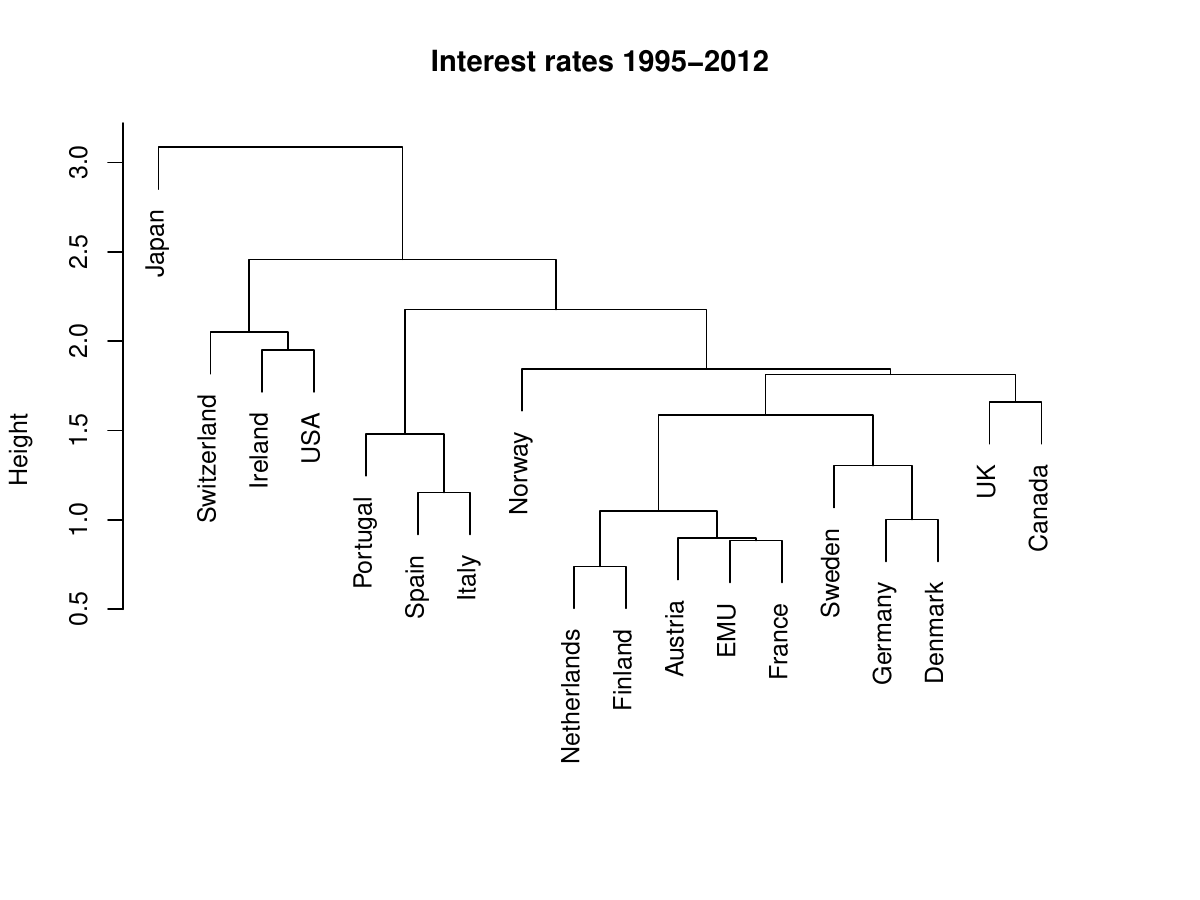}
\end{center}
\caption{Hierarchical clustering of several countries according to their interest rates from 1995 to 2012.}
\label{fig:interestrates_global}
\end{figure}

First, we define the \emph{returns} time series for each country and then we compute the corresponding WSD of any pair of countries (see Figure \ref{fig:interestwsd}). Next, we define the dissimilarity measure as the binary logarithm of the WSD mean plus $1$ (in order to avoid negative distances).
Finally, we plot the hierarchical clusters according to this dissimilarity measure (see Figure \ref{fig:interestrates_global}).

When defining the dissimilarity measure, we can restrict the WSD to only some areas instead of considering it entirely. For example, if we want to study the relationships between the different countries from the beginning of the century to the 2008 crisis at long-term scales, then we could only take into account the WSD area between 2001 and 2007, considering exclusively scales greater than 2 years. On the other hand, if border effects are relevant, only the WSD zone outside the cone of influence could be considered. However, in our example, border effects do not substantially alter the clustering result.

\begin{verbatim}
> library(wavScalogram)
> library(TSclust)
> data("interest.rates")
> returns <- apply(interest.rates, MARGIN = 2, function(x) diff(log(x)))
> Nsignals <- ncol(returns)
> countries <- colnames(returns)
> M <- Nsignals * (Nsignals - 1) / 2 # Number of pairings
> auxpair <- vector(mode = "list", M)
> k <- 1
> for (i in 1:(Nsignals - 1)) {
>   for (j in (i + 1):Nsignals) {
>     auxpair[[k]] <- c(i, j)
>     k <- k + 1
>   }
> }
> fwsd <- function(x) wsd(signal1 = returns[, x[1]],
+                         signal2 = returns[, x[2]],
+                         makefigure = FALSE)
> Allwsd <- lapply(auxpair, FUN = fwsd)
> ntimes <- length(Allwsd[[1]]$tcentral)
> nscales <- length(Allwsd[[1]]$scales)
> area <- ntimes * nscales
> meanwsd <- rep(0, M)
> for (i in 1:M) {
>   meanwsd[i] <- sum(Allwsd[[i]]$wsd) / area
> }
> d1 <- matrix(0, Nsignals, Nsignals)
> d1[lower.tri(d1, diag = FALSE)] <- log2(meanwsd + 1)
> dm1 <- as.dist(t(d1) + d1)
> names(dm1) <- countries
> plot(hclust(dm1), main = "Interest rates 1995-2012", xlab = "", sub = "")
\end{verbatim}

\subsection{Sunspots}

In the next example we are going to illustrate the different tools of \textbf{wavScalogram} on the most famous \emph{sunspot number time series} and how to use them in order to find the sunspots period, which is estimated  to be around $11$ years.

\begin{figure}[tbp]
\begin{center}
  \includegraphics[width=.95\textwidth]{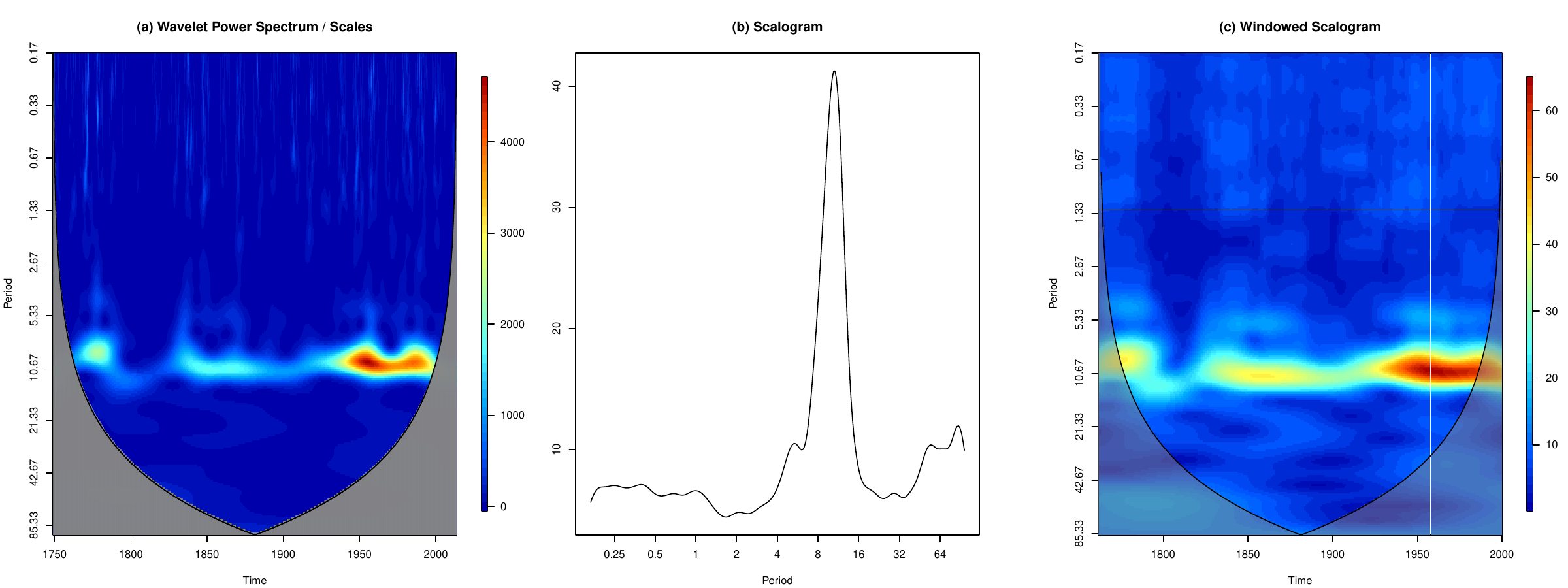}
\end{center}
\caption{(a) Wavelet power spectrum divided by scales, (b) scalogram, and (c) windowed scalogram of the sunspots time series, with \texttt{energy\_density = TRUE}. In plots (a) and (c) the yellow-red band should be centered in the sunspots period, while in plot (b), this period should be given by the peak in the scalogram.}
\label{fig:sunspots1}
\end{figure}

Let us consider the \texttt{sunspot.month} R dataset consisting on monthly numbers of sunspots from 1749 to present. Firstly, we can estimate the sunspots period by means of the scale at which the scalogram reaches its maximum. Using this criterion, we obtain that the sunspots period is $10.3254$ approximately (see Figure \ref{fig:sunspots1} (b)). For this method, it is recommended that \texttt{energy\_density = TRUE}\, since otherwise, larger scales would be over-estimated. Note that the wavelet power spectrum and the windowed scalograms present, as expected,  horizontal bands of high values precisely around the scale $10.3254$  (see Figure \ref{fig:sunspots1} (a) and (c)). Hence, they can be used to estimate a sunspot period that depends on time.

\begin{figure}[tbp]
\begin{center}
  \includegraphics[width=\textwidth]{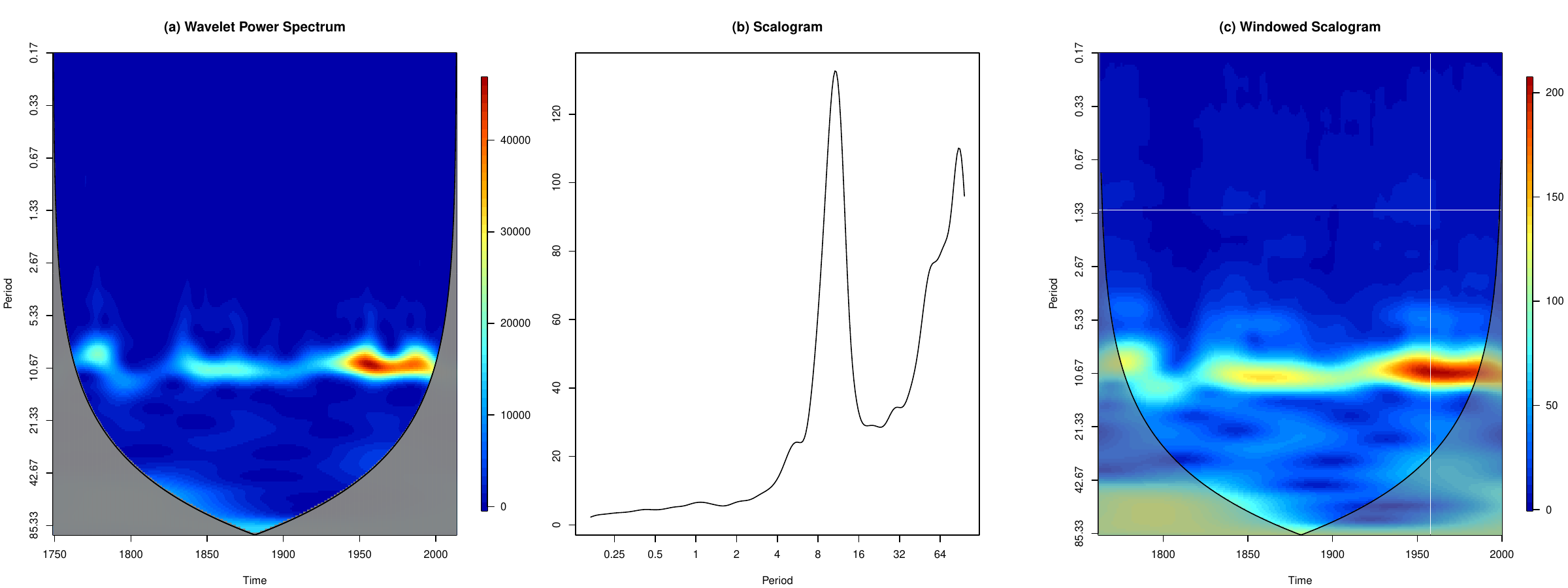}
\end{center}
\caption{(a) Wavelet power spectrum, (b) scalogram, and (c) windowed scalogram of the sunspots time series, with \texttt{energy\_density = FALSE}. These plots depict the same as Figure \ref{fig:sunspots1}, but the bias in favour of large scales is present. Nevertheless, this is recommended for computing the corresponding scale indices.}
\label{fig:sunspots2}
\end{figure}

\begin{figure}[tbp]
\begin{center}
  \includegraphics[width=\textwidth]{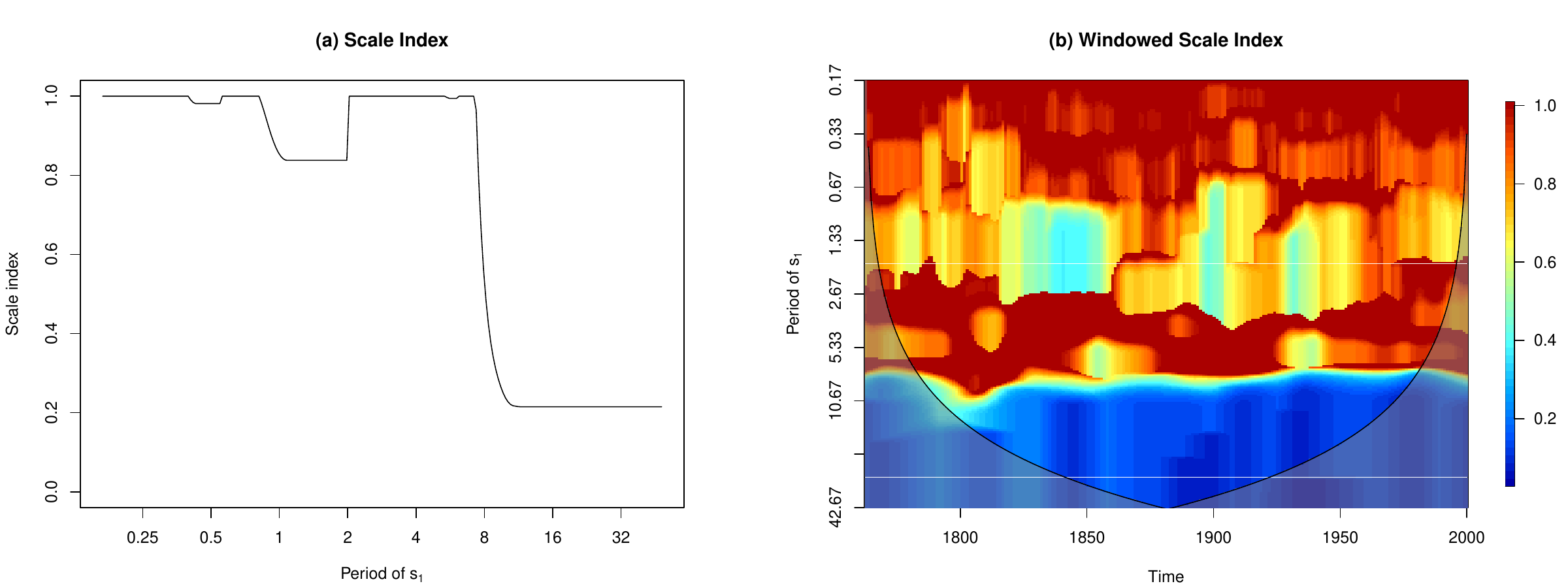}
\end{center}
\caption{(a) Scale indices and (b) windowed scale indices of the sunspots time series. In these plots, the use of the scale indices to determine the sunspots period is depicted. The period is estimated by the minimum scale $s_1$ for which the scale indices are stabilized around lower values, presenting the transition from non-periodicity to a far more periodic signal. The windowed scale indices in plot (b) are specially useful for non-stationary time series because they can detect changes in the sunspots period over time.}
\label{fig:sunspots3}
\end{figure}

On the other hand, we can also estimate the sunspots period by means of the scale at which the scale index reaches its minimum. Contrary to the previous case and according to Remark \ref{rmk:noed}, it is recommended that \texttt{energy\_density = FALSE} for computing scale indices (see Figure \ref{fig:sunspots2}). Using this criterion, we obtain that the sunspots period is $11.1215$, approximately (see Figure \ref{fig:sunspots3} (a)). Therefore, the windowed scale index can also be used, analogously to the scalogram and the windowed scalogram, to estimate sunspots periods depending on time (see Figure \ref{fig:sunspots3} (b)).

\bibliography{bolos-benitez}

\begin{thebibliography}{20}
\providecommand{\natexlab}[1]{#1}
\providecommand{\url}[1]{\texttt{#1}}
\expandafter\ifx\csname urlstyle\endcsname\relax
  \providecommand{\doi}[1]{doi: #1}\else
  \providecommand{\doi}{doi: \begingroup \urlstyle{rm}\Url}\fi

\bibitem[Aldrich(2020)]{wavelets}
E.~Aldrich.
\newblock \emph{wavelets: Functions for Computing Wavelet Filters, Wavelet
  Transforms and Multiresolution Analyses}, 2020.
\newblock URL \url{https://CRAN.R-project.org/package=wavelets}.
\newblock R package version 0.3-0.2.

\bibitem[Ben\'{\i}tez et~al.(2010)Ben\'{\i}tez, Bol\'os, and
  Ram\'{\i}rez]{ben10}
R.~Ben\'{\i}tez, V.~J. Bol\'os, and M.~E. Ram\'{\i}rez.
\newblock A wavelet-based tool for studying non-periodicity.
\newblock \emph{Computers \& Mathematics with Applications}, 60\penalty0
  (3):\penalty0 634--641, 2010.
\newblock URL \url{https://doi.org/10.1016/j.camwa.2010.05.010}.

\bibitem[Bol\'os and Ben\'{\i}tez(2021)]{wavScalogram}
V.~J. Bol\'os and R.~Ben\'{\i}tez.
\newblock \emph{wavScalogram: Wavelet Scalogram Tools for Time Series
  Analysis}, 2021.
\newblock URL \url{https://CRAN.R-project.org/package=wavScalogram}.
\newblock R package version 1.1.1.

\bibitem[Bol\'os et~al.(2017)Bol\'os, Ben\'{\i}tez, Ferrer, and Jammazi]{bol17}
V.~J. Bol\'os, R.~Ben\'{\i}tez, R.~Ferrer, and R.~Jammazi.
\newblock The windowed scalogram difference: a novel wavelet tool for comparing
  time series.
\newblock \emph{Applied Mathematics and Computation}, 312:\penalty0 49--65,
  2017.
\newblock URL \url{https://doi.org/10.1016/j.amc.2017.05.046}.

\bibitem[Bol\'os et~al.(2020)Bol\'os, Ben\'{\i}tez, and Ferrer]{bol20}
V.~J. Bol\'os, R.~Ben\'{\i}tez, and R.~Ferrer.
\newblock A new wavelet tool to quantify non-periodicity of non-stationary
  economic time series.
\newblock \emph{Mathematics}, 8:\penalty0 844--859, 2020.
\newblock URL \url{https://doi.org/10.3390/math8050844}.

\bibitem[Carmona and Torresani(2021)]{Rwave}
R.~Carmona and B.~Torresani.
\newblock \emph{Rwave: Time-Frequency Analysis of 1-D Signals}, 2021.
\newblock URL \url{https://CRAN.R-project.org/package=Rwave}.
\newblock R package version 2.6-0.

\bibitem[Daubechies(1992)]{dau92}
I.~Daubechies.
\newblock \emph{Ten Lectures on Wavelets}.
\newblock CBMS-NSF Regional Conference Series in Applied Mathematics. Society
  for Industrial and Applied Mathematics, 1992.
\newblock ISBN 9780898712742.

\bibitem[Farge(1992)]{far92}
M.~Farge.
\newblock Wavelet transforms and their applications to turbulence.
\newblock \emph{Annual Review of Fluid Mechanics}, 24:\penalty0 395--458, 1992.
\newblock URL \url{https://doi.org/10.1146/annurev.fl.24.010192.002143}.

\bibitem[Gouhier et~al.(2021)Gouhier, Grinsted, and Simko]{biwavelet}
T.~C. Gouhier, A.~Grinsted, and V.~Simko.
\newblock \emph{R package {biwavelet}: Conduct Univariate and Bivariate Wavelet
  Analyses}, 2021.
\newblock URL \url{https://github.com/tgouhier/biwavelet}.
\newblock (Version 0.20.21).

\bibitem[Liu et~al.(2007)Liu, San~Liang, and Weisberg]{liu07}
Y.~Liu, X.~San~Liang, and R.~H. Weisberg.
\newblock Rectification of the bias in the wavelet power spectrum.
\newblock \emph{Journal of Atmospheric and Oceanic Technology}, 24\penalty0
  (12):\penalty0 2093--2102, 2007.
\newblock URL \url{https://doi.org/10.1175/2007JTECHO511.1}.

\bibitem[Mallat(2008)]{mal98}
S.~Mallat.
\newblock \emph{A Wavelet Tour of Signal Processing: The Sparse Way}.
\newblock Academic Press, 2008.
\newblock ISBN 978-0-08-092202-7.

\bibitem[Montero and Vilar(2014)]{jss14}
P.~Montero and J.~Vilar.
\newblock {TSclust}: an {R} package for time series clustering.
\newblock \emph{Journal of Statistical Software}, 62\penalty0 (1):\penalty0
  1--43, 2014.
\newblock URL \url{https://doi.org/10.18637/jss.v062.i01}.

\bibitem[Nason(2022)]{wavethresh}
G.~Nason.
\newblock \emph{wavethresh: Wavelets Statistics and Transforms}, 2022.
\newblock URL \url{https://CRAN.R-project.org/package=wavethresh}.
\newblock R package version 4.6.9.

\bibitem[Roebuck and {Rice University's DSP group}(2022)]{rwt}
P.~Roebuck and {Rice University's DSP group}.
\newblock \emph{rwt: 'Rice Wavelet Toolbox' Wrapper}, 2022.
\newblock URL \url{https://CRAN.R-project.org/package=rwt}.
\newblock R package version 1.0.2.

\bibitem[Roesch and Schmidbauer(2018)]{WaveletComp}
A.~Roesch and H.~Schmidbauer.
\newblock \emph{WaveletComp: Computational Wavelet Analysis}, 2018.
\newblock URL \url{https://CRAN.R-project.org/package=WaveletComp}.
\newblock R package version 1.1.

\bibitem[Savchev and Nason(2018)]{hwwntest}
D.~Savchev and G.~Nason.
\newblock \emph{hwwntest: Tests of White Noise using Wavelets}, 2018.
\newblock URL \url{https://CRAN.R-project.org/package=hwwntest}.
\newblock R package version 1.3.1.

\bibitem[Taylor et~al.(2019)Taylor, Park, and Eckley]{mvLSW}
S.~A.~C. Taylor, T.~Park, and I.~A. Eckley.
\newblock Multivariate locally stationary wavelet analysis with the {mvLSW} {R}
  package.
\newblock \emph{Journal of Statistical Software}, 90\penalty0 (11):\penalty0
  1--19, 2019.
\newblock URL \url{https://doi.org/10.18637/jss.v090.i11}.

\bibitem[Torrence and Compo(1998)]{tor98}
C.~Torrence and G.~Compo.
\newblock A practical guide to wavelet analysis.
\newblock \emph{Bulletin of the American Meteorological Society}, 79\penalty0
  (1):\penalty0 61--78, 1998.
\newblock URL
  \url{https://doi.org/10.1175/1520-0477(1998)079<0061:APGTWA>2.0.CO;2}.

\bibitem[Torrence and Webster(1999)]{tor99}
C.~Torrence and P.~J. Webster.
\newblock Interdecadal changes in the {ENSO}–monsoon system.
\newblock \emph{Journal of Climate}, 12\penalty0 (8):\penalty0 2679--2690,
  1999.
\newblock URL
  \url{https://doi.org/10.1175/1520-0442(1999)012<2679:ICITEM>2.0.CO;2}.

\bibitem[Whitcher(2020)]{waveslim}
B.~Whitcher.
\newblock \emph{waveslim: Basic Wavelet Routines for One-, Two-, and
  Three-Dimensional Signal Processing}, 2020.
\newblock URL \url{https://CRAN.R-project.org/package=waveslim}.
\newblock R package version 1.8.2.

\end{thebibliography}
\bibliographystyle{abbrvnat}

\end{document}